\documentclass[aps,pre,showpacs,superscriptaddress,nofootinbib,12pt]{revtex4-1}

\usepackage{amsmath}
\usepackage{mathtools}
\usepackage{ulem}
\usepackage{graphicx}
\usepackage{color}

\usepackage{natbib}
\newcommand*{\chirdef}{N\left[ \left< \left< |v| \right>_t^2 \right>_s - \left<
\left< |v| \right>_t \right>_s^2 \right]}
\newcommand*{\chipsidef}{N\left[ \left< \left< |\psi| \right>_t^2 \right>_s -
\left< \left< |\psi| \right>_t \right>_s^2 \right]}

\DeclarePairedDelimiter\floor{\lfloor}{\rfloor}

\begin{document}


\title{Synchronization of Discrete Oscillators on Ring Lattices and Small-World
Networks}


\author{Kevin Liu Rodrigues}

\author{Ronald Dickman}
\affiliation{
Departamento de F\'{\i}sica, Instituto de
Ci\^encias Exatas, and National Institute of Science and Technology for Complex
Systems,\\
Universidade Federal de Minas Gerais \\
C.P. 702, 30123-970, Belo Horizonte, MG, Brazil
}%


\date{\today}

\begin{abstract}

A lattice of three-state stochastic phase-coupled oscillators introduced by
Wood {\it et al.\/} [Phys. Rev. Lett. {\bf 96}, 145701 (2006)] exhibits a phase
transition at a critical value of the coupling parameter $a$, leading to stable
global {\it oscillations} (GO). On a complete graph, upon further increase in
$a$, the model exhibits an {\it infinite-period} (IP) phase transition, at
which collective oscillations cease and discrete rotational ($C_3$) symmetry is
broken [Assis {\it et al.}, J. Stat. Mech.  (2011) P09023]. These authors
showed that the IP phase does not exist on finite-dimensional lattices.  In the
case of large negative values of the coupling, Escaff {\it et al.} [Phys. Rev.
E {\bf 90}, 052111 (2014)] discovered the stability of travelling-wave states
with no global synchronization but with local order. Here, we verify the IP
phase in systems with long-range coupling but of lower connectivity than a
complete graph and show that even for large {\it positive} coupling, the system
sometimes fails to reach global order. The ensuing travelling-wave state
appears to be a metastable configuration whose birth and decay (into the
previously described phases) are associated with the initial conditions and
fluctuations.

\end{abstract}


\maketitle


\section{Introduction}

Systems of coupled oscillators exhibit diverse symmetry-breaking transitions to
a globally synchronized state. In the paradigmatic Kuramoto model, for
instance, oscillators with distinct intrinsic frequencies $\omega_i$ coupled
via their continuous phases $\theta_i$ can exhibit stable collective
oscillations, breaking time-translation
invariance~\cite{Kuramoto84,Strogatz93,Strogatz00,StrogatzSync,Pikovsky01}.
Amongst discrete-phase models, the paper-scissors-stone game is an example of a
system with three absorbing states that can exhibit either global oscillations
or spontaneous breaking of discrete rotational ($C_3$)
symmetry~\cite{Tainaka88,Tainaka89,Tainaka91,Itoh94,Tainaka94}. More recently,
Wood and coworkers proposed a family of models of phase-coupled three-state
stochastic oscillators that undergo a phase transition to a state exhibiting
global oscillations (GO)~\cite{Wood06a,Wood06b,Wood07a,Wood07b} for
sufficiently strong coupling.  We shall refer to these as Wood's cyclic
model~(WCM). Although the WCM also has $C_3$ symmetry, it has no absorbing
state.  In addition to their intrinsic interest in the context of
nonequilibrium phase transitions, this family of models serve as a highly
simplified description of collective neuronal behavior.

The first WCM~\cite{Wood06a} was found to undergo a second phase transition
upon further increase of the coupling \cite{assis2011infinite}, at which the
period of oscillation becomes infinite, thereby breaking $C_3$ symmetry. In
\cite{assis2011infinite}, this infinite-period (IP) transition was studied on a
complete graph (all-to-all coupling), and a novel order parameter, involving
the mean rate of change of the probability distribution, was proposed. On the
basis of a nucleation scenario, the authors of Ref. \cite{assis2011infinite}
argued {\it against} the existence of an IP transition on finite-dimensional
lattices with short-range interactions, but left open the question of its
occurrence on networks with nonlocal interactions. Here we study a WCM on (1)
regular rings with interactions up to $2K$-neighbors, varying the interaction
range $K$, and (2) small-world networks. Using numerical simulations to study
the order parameter and its variance, we verify the existence of GO and IP
phase transitions on these structures. For regular rings of $N$ nodes, the
degree of connectivity is characterized by $\alpha \equiv K/N$ such that
$\alpha \in (0, 0.5)$.  A key question is the minimum value of $\alpha$
necessary to observe the GO and IP phases as $N \to \infty$.  Our results
suggest that any $\alpha > 0$ is sufficient.

We also provide evidence that for intermediate interaction ranges, global
synchronization depends sensitively on initial conditions: Some realizations
with a random initial configuration show no global synchronization. Such events
persist even as the number of neighbors grows in proportion to the system size.
In this situation, the final state may be a travelling wave, as observed by
Escaff et al. \cite{escaff2014arrays} in the case of {\it anti-crowding}, i.e.,
interactions favoring anti-synchronization between neighbors.

The remainder of this paper is organized as follows. In section~\ref{model} we
review the WCM and the essentials of the transitions to the GO and IP phases.
We report our results on the GO and IP phase transitions on regular rings, and
on small-world networks, in  sections \ref{regularrings} and \ref{smallworld},
respectively.  Our conclusions are discussed in section~\ref{conclusions}.

\section{\label{model} Model}

\begin{figure}[!b]
\begin{center}
\fbox{\includegraphics[width=0.35\textwidth]{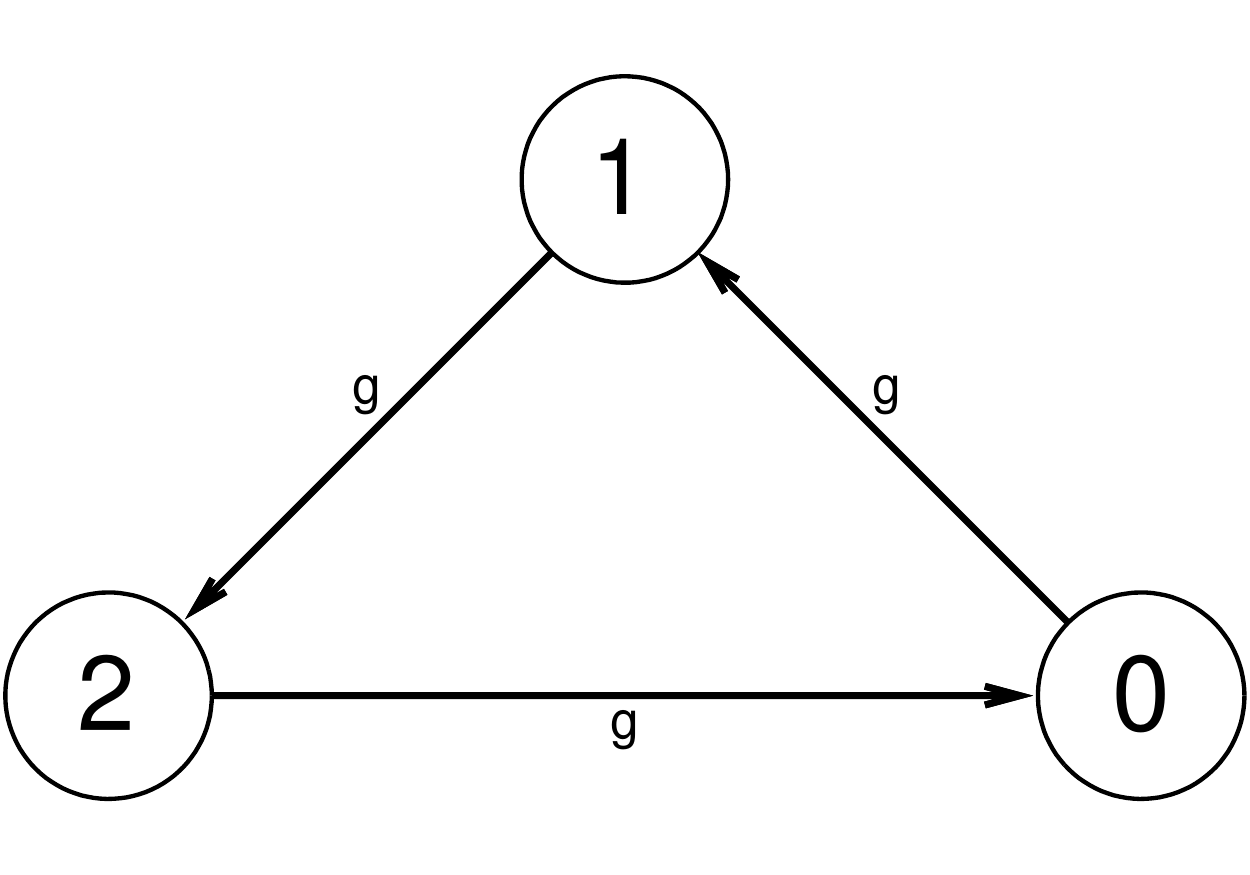}}
\caption{\label{fig:taxas}
    Transition rates for an isolated unit.
    }
\end{center}
\end{figure}

\begin{figure}
\includegraphics[height=.75\textheight]{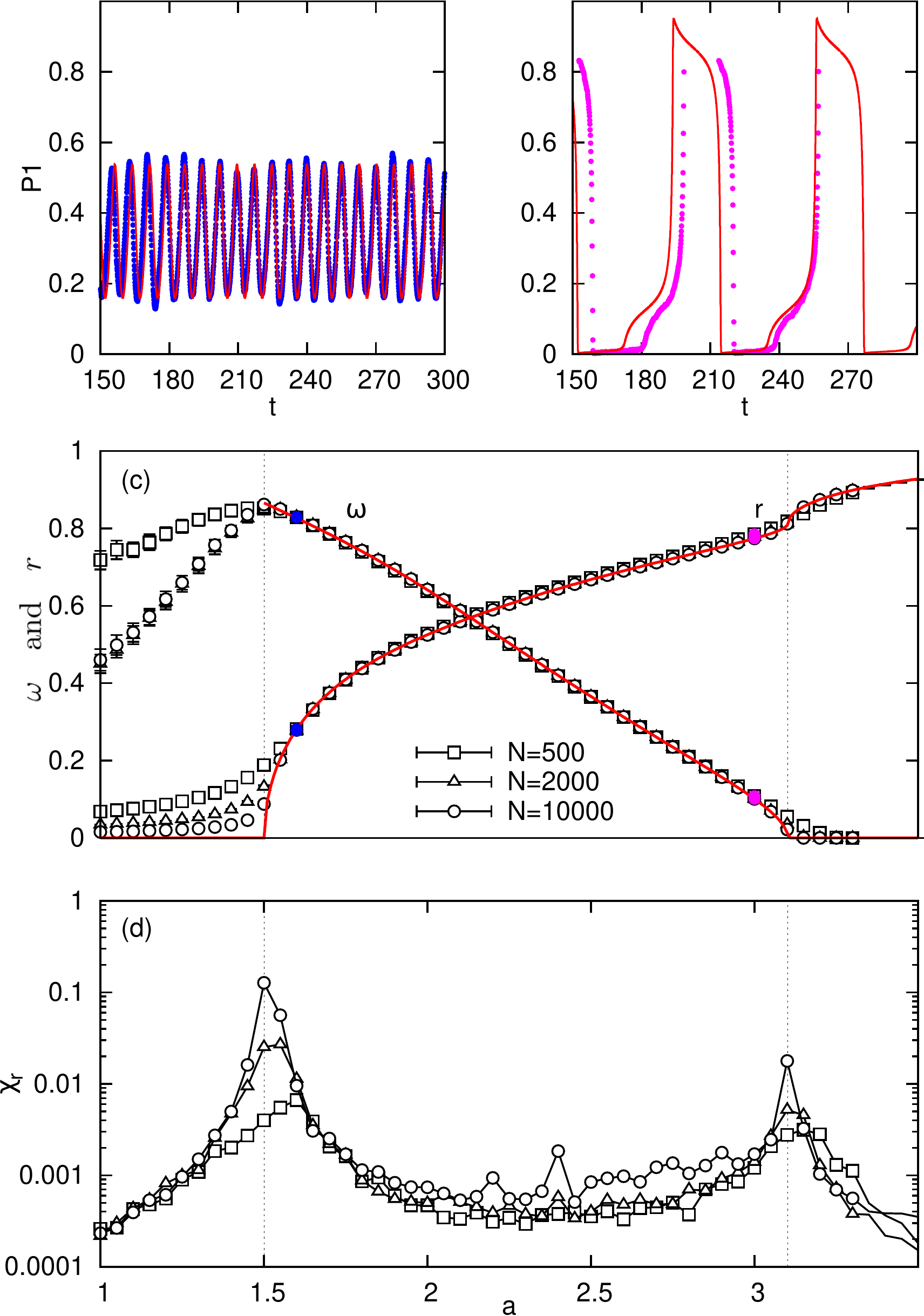}
\begin{center}
\caption{\label{complete_graph}
    (Color online) Panels (a) and (b) show the evolution of $P_1$ for $a=1.6$
    and $a=3$, respectively. Points: simulations of a complete graph of $N$
    nodes; lines: mean-field solution. (c) Dependence of $r$ and $\omega$ on
    $a$, exhibiting the two phase transitions. (d) $\chi_r$ versus $a$, showing
    peaks at the transitions [system sizes as in (c)].
    }
\end{center}
\end{figure}

In the WCM, the state $j_x$ at site $x$ ($x=1,\ldots,N$) can take one of three
values, $j_x \in \{0,1,2\}$, corresponding to a phase $\phi_x = 2 \pi j_x/3$.
The only allowed transitions are those from $j$ to $j+1$ (modulo 3) (see
Fig.~\ref{fig:taxas}), which implies that detailed balance is violated. If site
$x$ is in state $j$, its transition rate to state $j+1$ is:

\begin{equation}
\label{eq:gj}
g_{j, j+1} = g \exp \left[ \frac {a \left( n_{x,j+1} - n_{x,j} \right)} {z}
    \right]
\end{equation}

\noindent where $g$ is a constant rate, $a$ is the coupling parameter, $n_{x,k}$
is the number of nearest neighbors of site $x$ in state $k$, and $z$ is the
number of nearest neighbors. Since these rates are invariant under cyclic
permutation of the state indices, the model is invariant under the group $C_3$
of discrete rotations.

Let $N_j$ be the total number of sites in state $j$, so that $N_0 +N_1 +N_2 =
N$, the total number of sites. As discussed in \cite{assis2011infinite}, the
mean-field (MF) approximation, obtained by replacing $n_{x,j}/z$ in the argument
of the exponential of~Eq.(\ref{eq:gj}) with the corresponding state fraction,
$n_j = N_j/N$, yields three nonequilibrium phases, separated by two continuous
phase transitions. For small coupling ($a < a_c = 1.5$), the disordered phase,
with ${\bf n}=(1/3,1/3,1/3) \equiv {\bf n}_{1/3}$, is the stable stationary
solution of the MF equations. (${\bf n}$ denotes the vector of state fractions.)
For $a$ between $a_c$ and a higher value, $a^c\simeq 3.102\, 439\, 915\, 64$,
there is no stable stationary solution and the MF equations admit an
oscillatory solution (a limit cycle) in which states 0, 1 and 2 periodically
assume the role of the majority. As $a$ is increased above $a_c$, the frequency
$\omega$ of oscillation decreases continuously, becoming zero at $a^c$,
signalling the IP transition.  For $a> a^c$, three stationary solutions, {\bf
n}$_i$ appear, such that state $i$ represents the (permanent) majority. Thus
$C_3$ symmetry is broken for $a > a^c$. (The three solutions are, naturally,
related via cyclic permutation of indices in state space.) 

The WCM is characterized by a pair of order parameters. First, one has the
familiar Kuramoto synchronization parameter
~\cite{Kuramoto84,Strogatz00,Wood06a},

\begin{equation}
    r \equiv \left<\left< |v| \right>_t\right>_s,
    \label{eq:r}
\end{equation}
\noindent where
\begin{equation}
    v \equiv \frac{1}{N} \sum^{N}_{x=1} e^{i\phi_x}.
    \label{eq:v}
\end{equation} 

\noindent In Eq.~(\ref{eq:r}),  $\left< \right>_t$ denotes a time average over a
single realization (in the stationary state), and $\left< \right>_s$ an average
over independent realizations. Note that $r > 0$ is consistent with, but does
not necessarily imply, globally synchronized oscillation. The latter is
characterized by a periodically varying phase of
$v$~\cite{Kuramoto92a,Ohta08,ShinomotoKuramoto86,Rozenblit11}.

In the MF analysis, the transition to the synchronized regime (the GO
transition) is associated with a supercritical Hopf bifurcation at $a=a_c=1.5$:
the trivial fixed point $\bf{n}_{1/3}$ loses stability at $a=a_c$, and a limit
cycle encircling this point appears. For $a \gtrsim a_c$, sustained oscillations
in $n_j$ characterize synchronization among the oscillators
(Fig.~\ref{complete_graph}a).  Correspondingly, $r$ grows continuously $\sim
(a-a_c)^\beta$ at the transition (Fig.~\ref{complete_graph}c), with a mean-field
exponent $\beta = 1/2$~\cite{Wood06a}. The scaled variance

\begin{equation}
    \chi_r \equiv L^d \left[ \left< \left< |v| \right>^2_t \right>_s - r^2
    \right],
    \label{eq:chir}
\end{equation}

\noindent diverges with the system size at criticality, as shown in
Fig.~\ref{complete_graph}d for simulations on the complete graph
\cite{assis2011infinite}. The GO transition is associated with breaking of the
continuous time-translation symmetry: the $n_j(t)$, are periodic for $a\gtrsim
a_c$.  Increasing $a$ above $a_c$ enhances synchronization among the
oscillators, leading to increasing oscillation amplitudes, as shown in
Fig.~\ref{complete_graph}b.

Wood {\it et al.} found that the increasing amplitude of oscillation is
accompanied by a decreasing angular frequency $\omega = 2\pi/\langle \tau
\rangle$, where $\langle \tau \rangle$ is the mean time between peaks in $n_k$
(Figs.~\ref{complete_graph}a-c). This can be understood qualitatively from the
exponential dependence of the transition rates of Eq.~(\ref{eq:gj}) on the
neighbor fractions: When a state is highly populated, the rate at which
oscillators leave it becomes very small.  In mean-field theory, when $a$ reaches
the upper critical value $a^c$, three symmetric saddle-node bifurcations occur
simultaneously, and the period of the collective oscillations diverges
\cite{assis2011infinite}. Above $a^c$, there are three symmetric attractors in
the system, and 3-fold rotational ($C_3$) symmetry is spontaneously broken.  As
in condensation or a ferromagnetic phase transition~\cite{Huang}, freezing of
the majority state does not imply that individual oscillators freeze as well.
The transition rates of individual oscillators do decrease with increasing $a$,
but only vanish in the limit $a\to\infty$, when one of the states is fully
occupied.

It is convenient to define an order parameter $\psi$ that is identically zero
(in the infinite-size limit) for $a>a^c$. Assis et al. proposed
\cite{assis2011infinite},
\vspace{1em}

\begin{equation}
    \label{eq:psi}
    |\psi| \equiv \frac{1}{N}\left| \sum_{x=1}^N \left( \delta_{0,j_x} +
    e^{2\pi i/3}\delta_{1,j_x} + e^{-2\pi i/3}\delta_{2,j_x} \right) \gamma_x
    \right|,
\end{equation}
\vspace{1em}

\noindent where $\delta_{ij}$ is the Kronecker delta and $\gamma_x \equiv
g_{j_x,j_x +1}$ is the transition rate at site $x$ (see eq.~\ref{eq:gj}). Thus
$|\psi|$ involves not only the configuration, but the rate at which the latter
evolves.  On the complete graph, $\gamma_x$ is the same for all sites $x$ in the
same state $j$. Denoting this rate by $\gamma_j$, the order parameter can be
written (in MF analysis) as

\begin{equation}
\label{eq:psiMF}
|\psi|^2 \stackrel{MF}{=} \sum_{j=0}^2 (n_j \gamma_j - n_{j-1} \gamma_{j-1})^2 
\end{equation}

\noindent Both the disordered phase ($a< a_c$) as well as the IP phase
($a>a^c$) have stable, stationary solutions, ${\bf n}^*$. Since $\dot{n}_j = 0$
implies $n^*_j\gamma_j = n^*_{j-1}\gamma_{j-1}$, i.e., zero net change in the
probability of state $j$, we have $|\psi|=0$ in eq.~\ref{eq:psiMF} for both
cases [a similar line of reasoning can be applied directly to
Eq.~(\ref{eq:psi})].

\begin{figure}
\begin{center}
\includegraphics[width=0.7\textwidth]{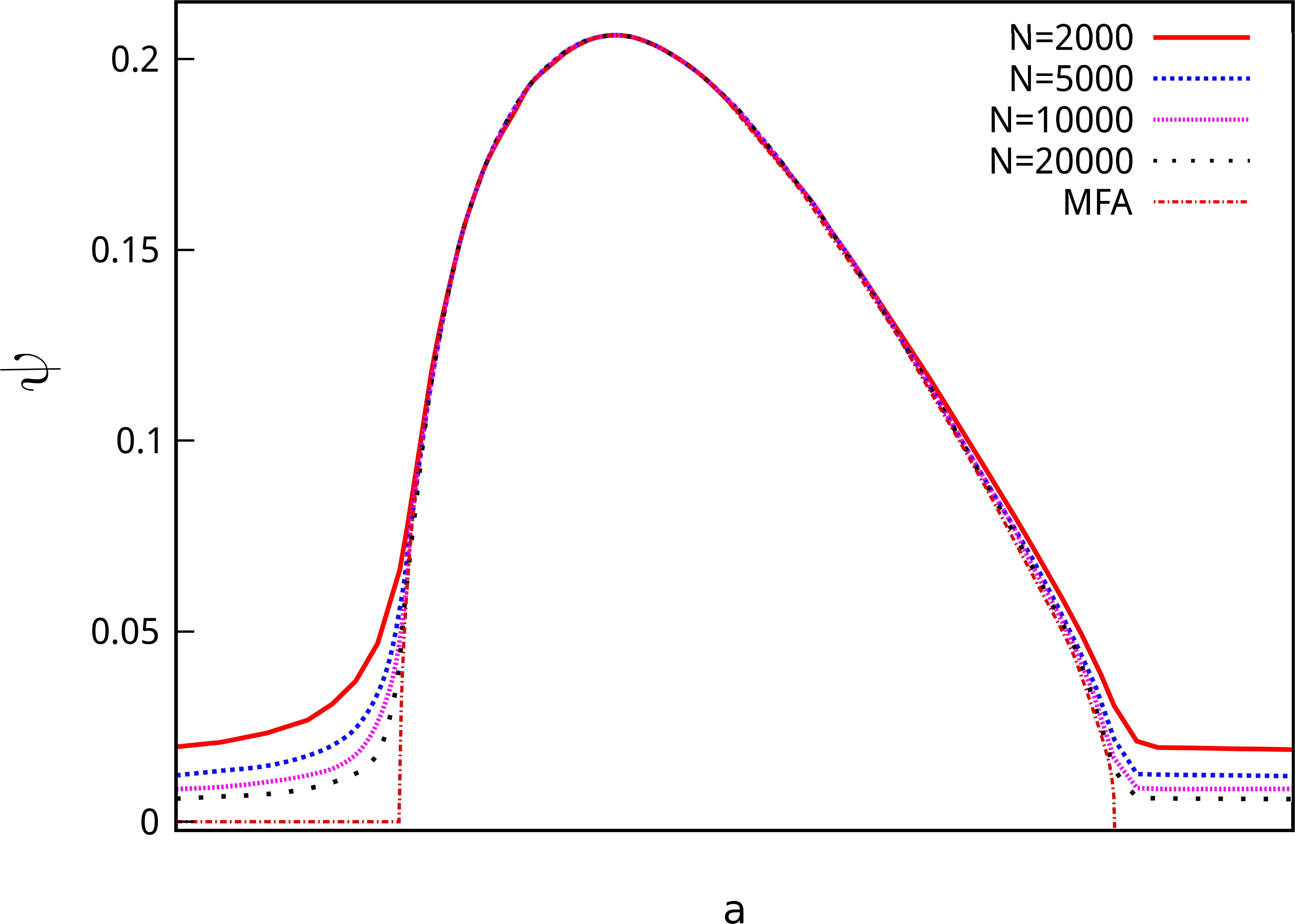}
    \caption{\label{fig:psi}
        (Color online) (From \citep{assis2011infinite}.) Order parameter $\langle
        |\psi| \rangle$ as a function of coupling $a$ in mean-field theory and
        on the complete graph, for sizes as indicated.
    }
\end{center}
\end{figure}

Figure \ref{fig:psi} shows $\langle |\psi| \rangle$ versus $a$ in MF theory, and
on the complete graph (the latter via numerically exact solution of the master
equation), confirming that $|\psi|$ functions as an order parameter to detect
both the GO and IP phase transitions. The MF critical behavior is $\langle
|\psi| \rangle \sim (a- a_c)^{1/2} $ for $a \searrow a_c$ and $\langle |\psi|
\rangle \sim (a^c - a)^{1/2}$ for $a \nearrow a^c$.  On the complete graph, the
order parameter decays with system size as $\langle |\psi| \rangle \sim
N^{-1/4}$ at $a=a_c$ and as $N^{-0.4203(3)}$ at $a=a^c$. The first result is
typical of mean-field-like scaling with system size at a continuous phase
transition, as argued in \cite{assis2011infinite}.

The results for the IP transition in MF and on a complete graph are in sharp
contrast to what is found on finite-dimensional lattices. The {\it absence} of
such a transition was verified numerically on hypercubic lattices in dimensions
$d \leq 4$ in \cite{assis2011infinite}. This reference also provides a
quantitative argument showing that on finite-dimensional lattices, a $j$-state
majority cannot persist indefinitely: it is always susceptible to change via
nucleation of a cluster of state $j+1$. The authors of \cite{assis2011infinite}
conjectured that the IP transition would occur on structures in which a site
interacts with a nonzero fraction of all other sites (as the system size tends
to infinity). In the following sections we test this conjecture on two
structures, regular rings with extended interactions, and small-world networks.

\section{\label{regularrings} The WCM on Regular Rings }

\begin{figure}[b]
\begin{center}
\fbox{\includegraphics[width=0.40\textwidth]{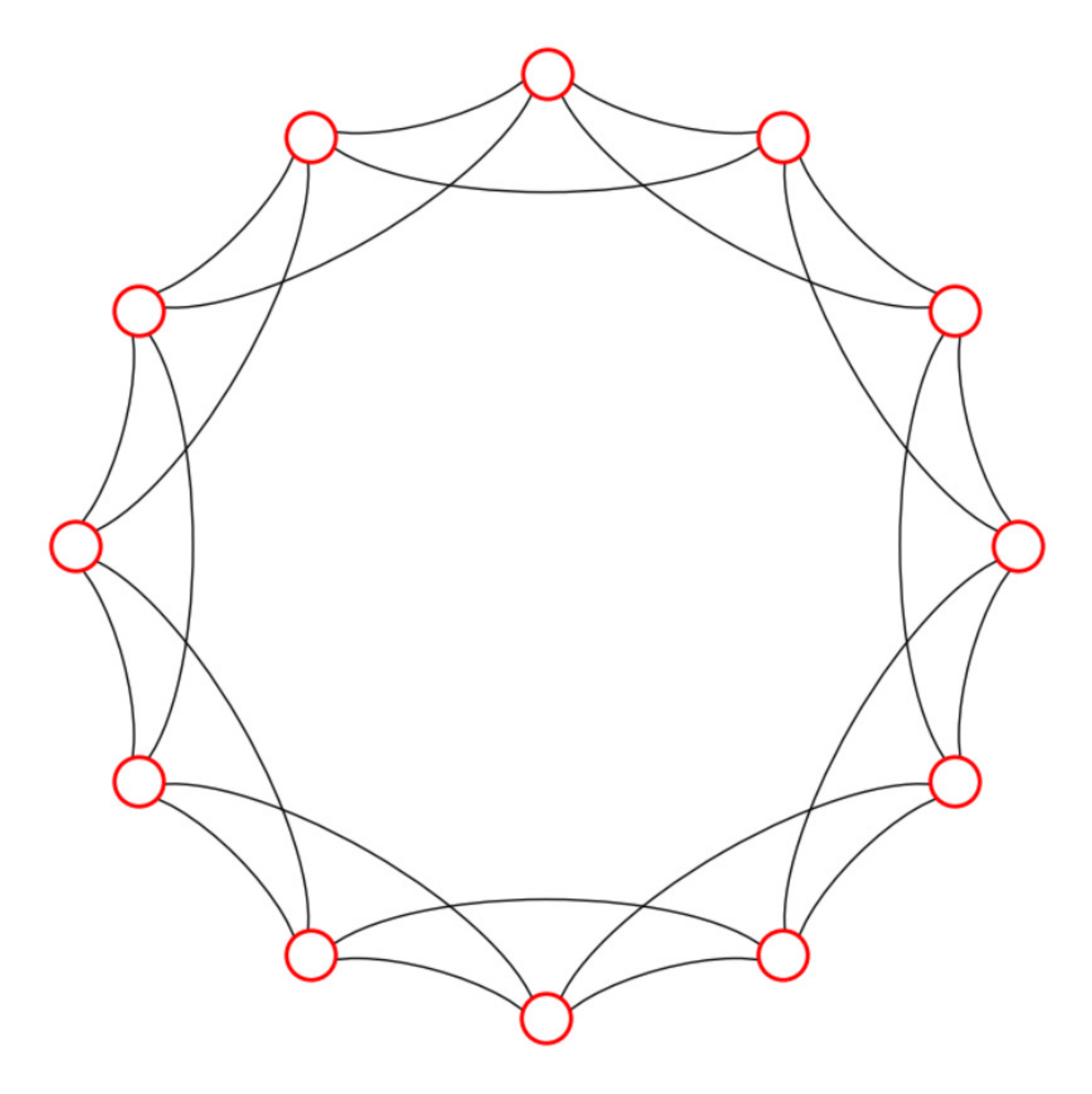}}
\caption{\label{fig:ring}
    A regular ring is an undirected graph with $N$ nodes arranged in circular
    fashion, with each connected to its $K$ nearest neighbors in each direction.
    Here we show an example with $N=12$ and $K=2$.
    }
\end{center}
\end{figure}

A {\it regular ring} is constructed starting from a graph of $N$ nodes arranged
in circular fashion. Considering one node at a time in a clockwise manner, we
connect it to its $K$ nearest neighbors in the clockwise direction; an example
of such a structure is shown in Fig.~\ref{fig:ring}. We define the {\it
connectivity} of a regular ring graph by $\alpha \equiv K/N$, such that
$\alpha=1/N$ signifies a one-dimensional chain while $\alpha=0.5$ represents a
complete graph. Thus, one can interpolate from the one-dimensional to an
infinite-dimensional hypercubic lattice (complete graph) varying $\alpha$ over
the interval $\left( 0, \frac{1}{2} \right]$.  It is known that the WCM on
hypercubic lattices of dimensions 1 and 2 cannot sustain ordered phases
\cite{Wood06a, assis2011infinite}. Since $\alpha=0.5$ represents the complete
graph, there must be at least one threshold $\alpha=\alpha^*$ above which one or
both phase transitions (GO and IP) occur, varying $a$.

Different from hypercubic lattices, in which coupling is local, for 
$\alpha > 0$ the coupling on ring graphs is nonlocal.  The manner
in which the interaction range scales as $N\to\infty$ can be chosen in different
ways; the simplest, which we consider here, is to fix $\alpha$ so that $K
\propto N$ (More precisely,  $K=\lceil\alpha N\rceil$ where $\lceil x \rceil$
denotes the smallest integer larger than $x$).  It is reasonable to
expect that phase transitions occur for any fixed $\alpha > 0$, since the
interaction range $K$ then tends to infinity with $N$. 

\begin{figure}[t]
\begin{center}
\includegraphics[width=0.9\textwidth]{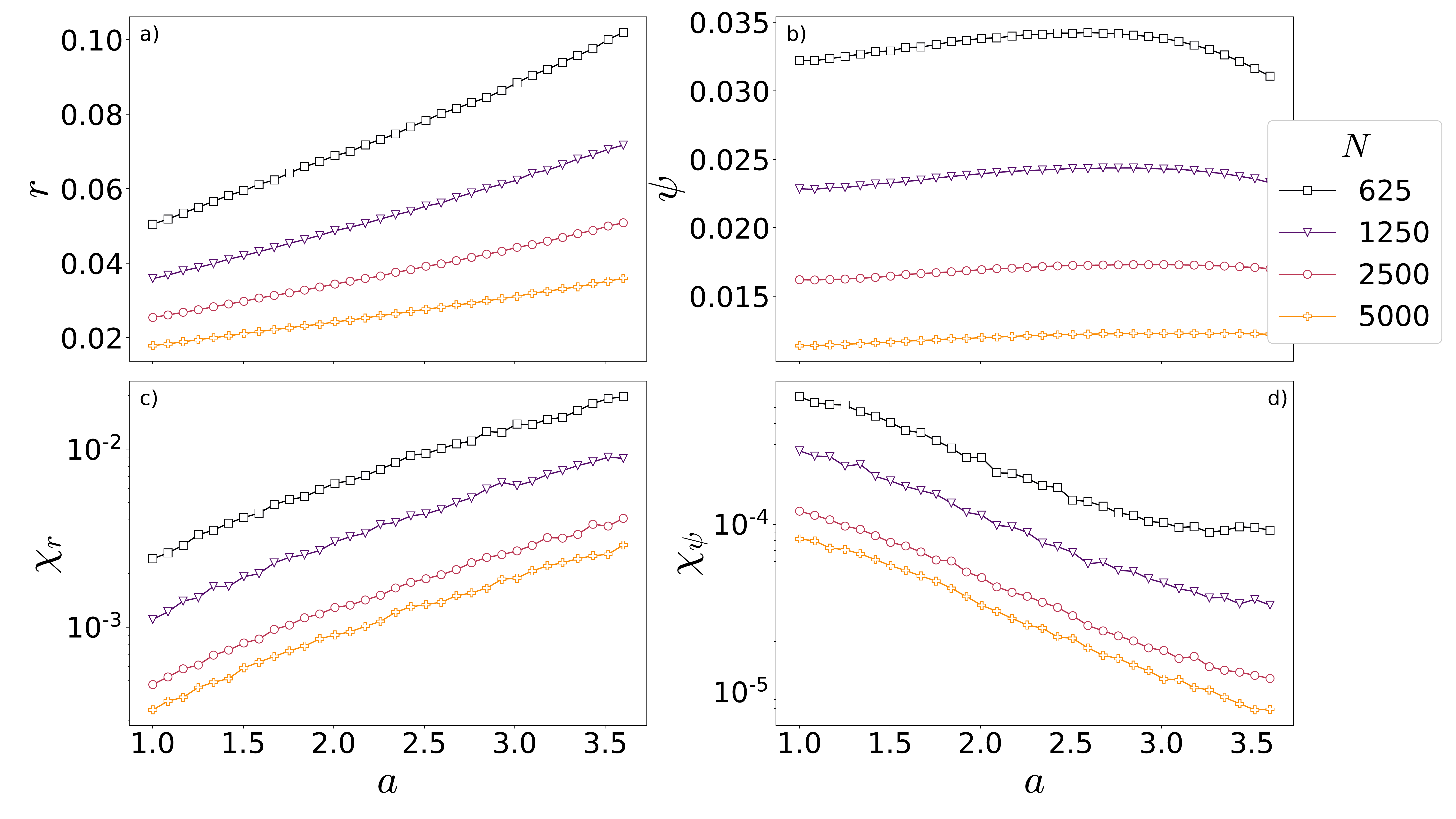}
\caption{\label{fig:chi_curves_1D}
    (Color online) Order parameters and their scaled variances for
    one-dimensional rings ($K=1$). Both the order parameters and their
    respective variances decrease as the system size is increased, with $r
    \approx \psi \approx 0$ across a wide range of coupling strengths. Points
    represent an average over 3000 independent realizations with random initial
    configurations. For the 1D chain, the same behavior is observed regardless
    of initial configuration.
    }
\end{center}
\end{figure}

\begin{figure}[t]
\begin{center}
\includegraphics[width=0.95\textwidth]{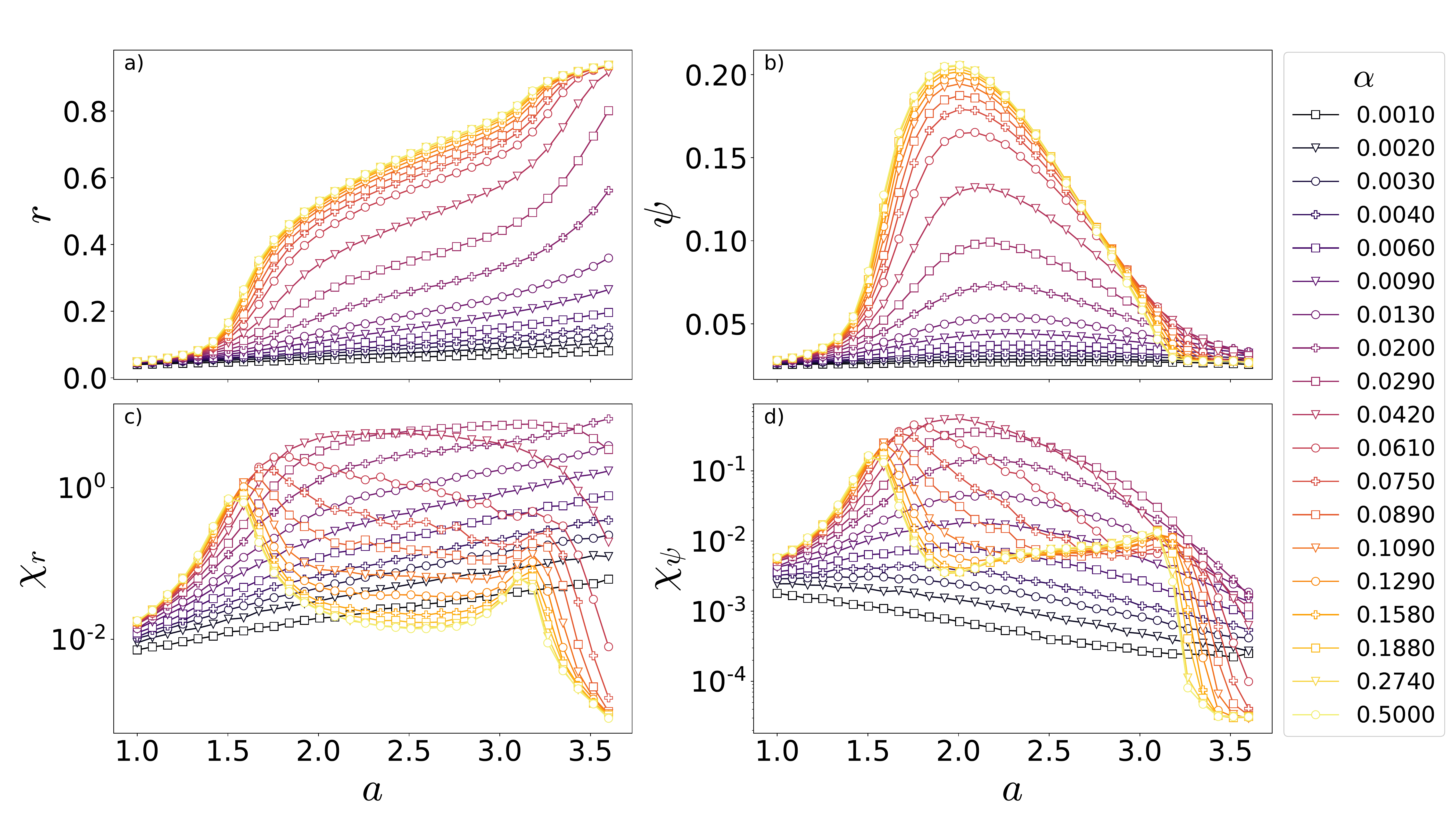}
    \caption{\label{fig:chicurves}
    (Color online) Order parameters and scaled variances for regular rings of
    size $N=1000$ and various values of the connectivity $\alpha$.  Points
    represent an average over 4000 independent realizations with uniform initial
    configurations.
    }
\end{center}
\end{figure}

\subsection{Scaling behavior: phase boundary}

The dynamics as $N$ increases can be studied by defining the scaled variances of
the order parameters. These quantities are expected to diverge in the
thermodynamic limit when the system undergoes a continuous phase transition
\cite{plischke1994equilibrium}.  The scaled variances of order parameters $r$
and $\psi$ can be defined through equations \ref{eq:v} and \ref{eq:psi} as:

\begin{align}
    \chi_r &= \chirdef \notag , \\
    \chi_\psi &= \chipsidef,
\end{align}

\noindent where $\left<\right>_t$ and $\left<\right>_s$ are averages over time
and over independent realizations, respectively.

In simulations, the system is allowed to relax to a steady state, starting from
its initial configuration.  Once the steady state has been attained, the order
parameters are averaged over the remainder of the evolution. As we will see, the
choice of initial condition is important for some values of the interaction
range $K$; we will focus on two different setups: \textit{random} initial
configurations, in which the initial phase of each oscillator is chosen
uniformly and independently among the three possible values in $\{ 0, 2\pi/3,
4\pi/3 \}$, and a \textit{uniform} initial condition, in which $\phi_i = 0, \;
\forall i$.

\begin{figure}
\begin{center}
\includegraphics[width=0.9\textwidth]{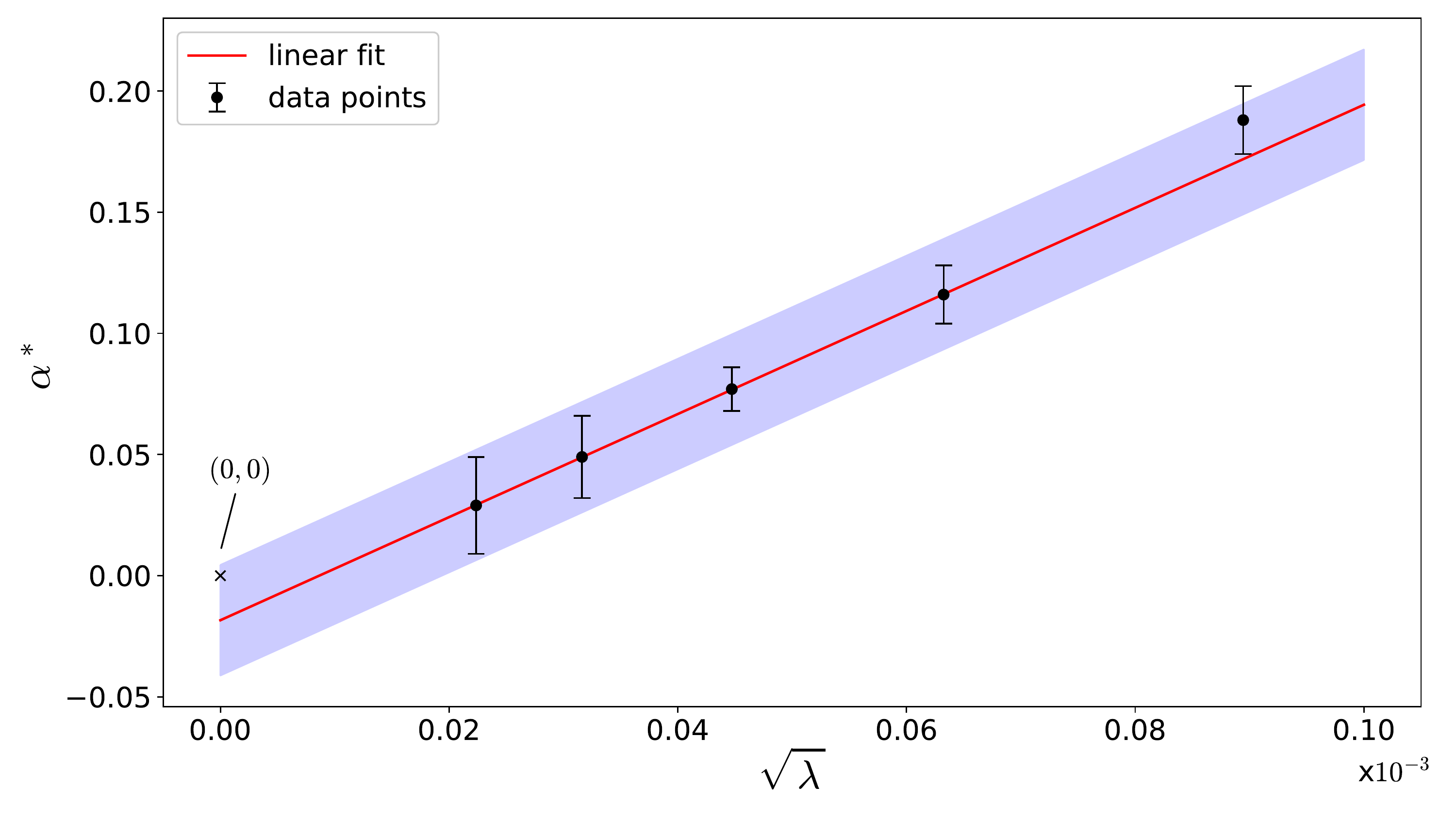}
    \caption{\label{fig:alphasplit} 
        Plot of $\alpha^*$ versus $\sqrt{\lambda}$. Starting from uniform
        initial configurations, $\alpha > \alpha^*$ means $\chi_{\psi}$ and
        $\chi_r$ exhibit two maxima, whereas for $\alpha < \alpha^*$ only a
        single broad maximum is observed. Full circles represent data obtained
        from simulations and solid curve is a linear fit with equation
        $y=-0.01837+2.1271x$. The band around the fitted curve represent the
        uncertainty associated with the linear intercept.}
\end{center}
\end{figure}

In the following we describe results for uniform initial conditions. In
Fig.~\ref{fig:chi_curves_1D} we plot the order parameters and their associated
variances for $K=1$ (i.e.,  $\alpha=1/N$). Both quantities are shown to
decrease with system size (for the 1D case in particular this behavior is the
same regardless of initial configuration), indicating the absence of phase
transitions. In Fig.~\ref{fig:chicurves}, the same quantities are shown for
regular rings with $N=1000$ and various $\alpha$ values.  As expected, the
order parameters and their variances approach the complete-graph limit as
$\alpha$ nears the value $1/2$.  Denoting by $\alpha^*$ the value associated
with a change from one to two maxima in $\chi_{\psi}$, we identify $\alpha^*
\approx 0.06$ for $N=1000$ in Fig.~\ref{fig:chicurves}.  Performing similar
analyses for different system sizes we obtain $\alpha^*$ as a function of $N$.
To infer the scaling behavior, we define $\lambda \equiv N^{-1}$ and look at
the $\alpha^*$ versus $\sqrt{\lambda}$ curve near $\lambda=0$. The resulting
data, shown in Fig.~\ref{fig:alphasplit}, suggest that $\alpha^*$ tends to zero
as $N\to\infty$. This supports the conjecture stated previously that in this
limit, and for any fixed $\alpha>0$, the WCM on a regular ring lattice exhibits
both GO and IP phase transitions.

\subsection{Scaling behavior: order parameter}

To better understand the scaling behavior we look at the order parameters as $N
\to \infty$ with fixed $\alpha$. If both $r$ and $\psi$ tend to zero, there is
no global synchronization. Both order parameters tending to positive values
indicates the presence of global or intermittent synchronization among large
populations of oscillators, while $\psi \to 0$ with $r \sim 1$ defines an
infinite-period phase.

\begin{figure}
\begin{center}
    \includegraphics[width=0.85\linewidth]{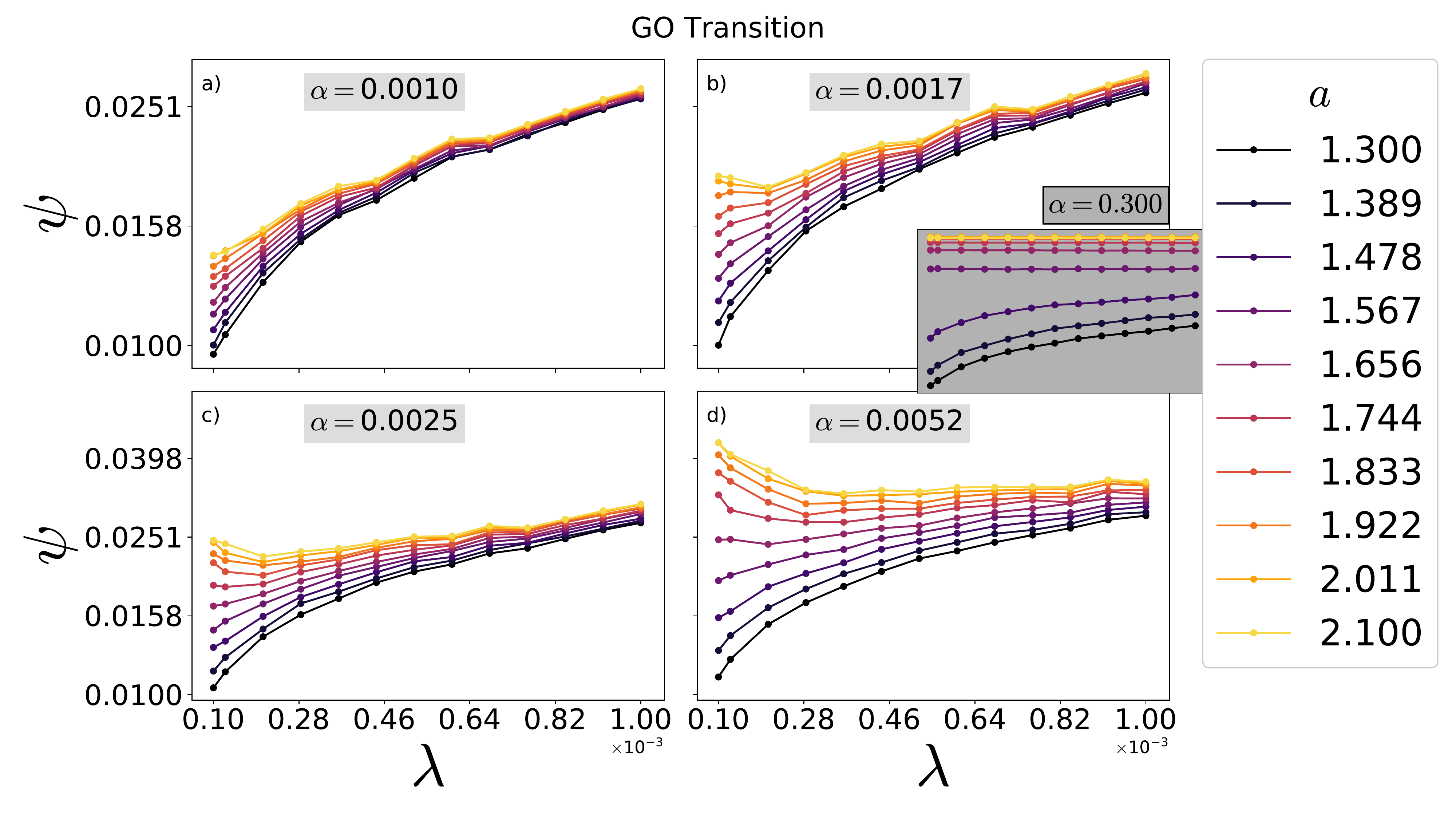}
    \includegraphics[width=0.85\linewidth]{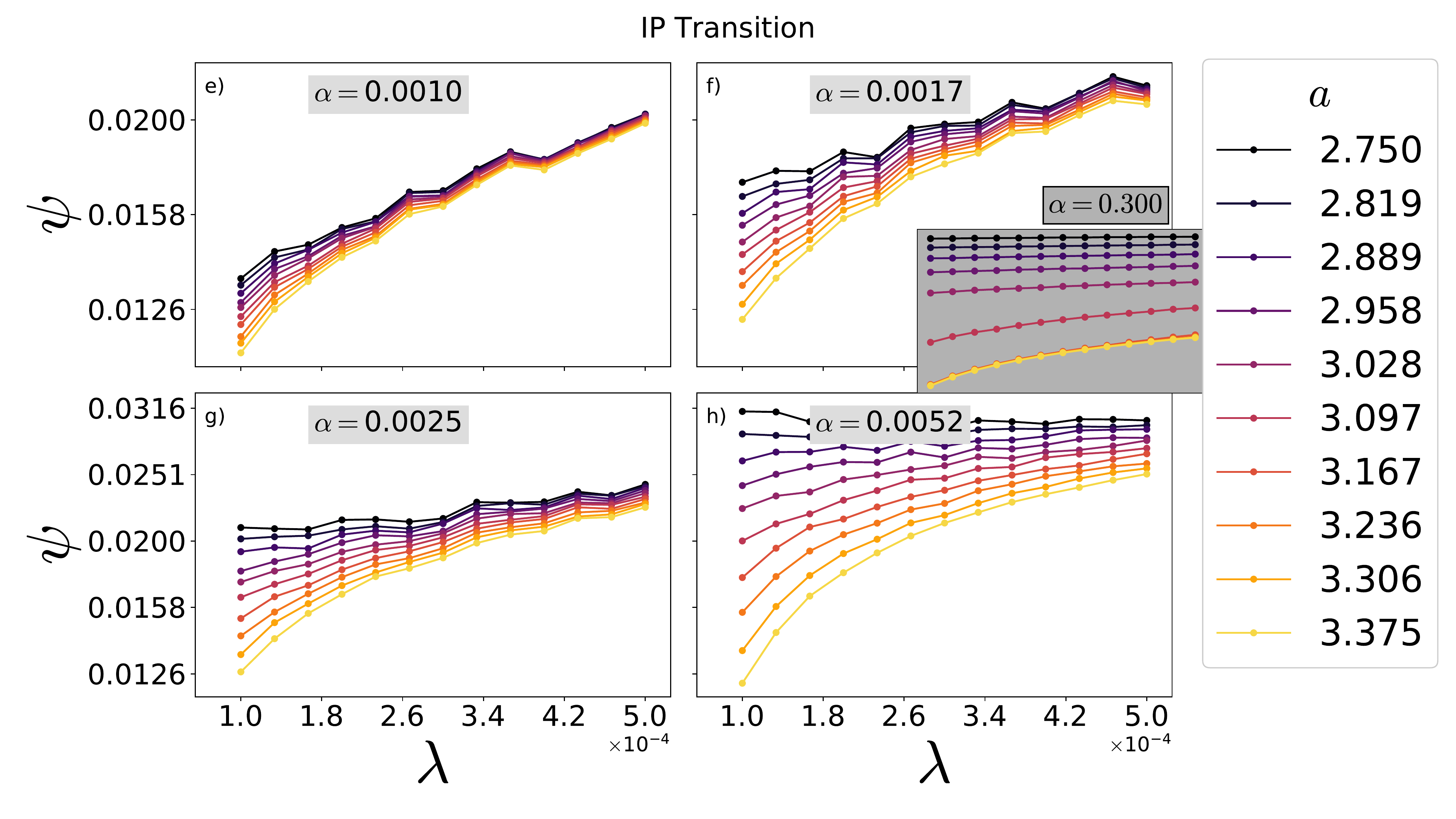}
\end{center}
\caption{\label{fig:opsplit}
    (Color online) Order parameter $\psi$ versus inverse system size $\lambda$
    for various values of $\alpha$, and coupling strengths $a$ in the vicinity
    of the GO and IP phase transitions. The insets in \textbf{b} and \textbf{f}
    show the behavior for $\alpha=0.3$, approaching the complete graph, with GO
    and IP transitions near $a_c=1.5$ and $a^c\approx3.1$ respectively.  Each
    point represents an average over 400 independent realizations.  The
    oscillatory behavior seen in some panels is due to the reduction in average
    path lengths caused by the introduction of new neighbors as the system size
    increases at fixed $\alpha$, and the fact that $K$ must be an integer (see
    Appendix \ref{appendix:LC}).}
\end{figure}

In this context it is useful to plot the order parameter versus
$\lambda\equiv1/N$. An upward (downward) curvature as $\lambda \to 0$ signals a
nonzero (zero) value of the order parameter. In Fig.~\ref{fig:opsplit}, such
plots are shown for selected values of $\alpha$ \footnote{{See full animations
        of Fig.~\ref{fig:opsplit}:
\href{https://youtu.be/pPQbc0eiv_4}{GO transition}},
\href{https://youtu.be/_qfNzoBpRO4}{IP transition}}
and system sizes up to $N=10^4$.  In panel b) we see evidence of the GO
transition for the value $\alpha=0.0017$ in the form of an inversion in
curvatures for lines of constant coupling strength, which happens at $a_c
\approx 2$.  The inset in this same panel shows that for large $\alpha$ there
is a clear split near the complete graph value $a_c=1.5$.  In the case of the
IP transition we observe that there is no upward curvature at any point, but
rather a sharp increase in density of the lines of constant $a$ for higher
values of $\alpha$, as seen in the bottom inset of Fig.~\ref{fig:opsplit} where
the lines $a=3.375$ and $a=3.306$ have collapsed to the same point near the
origin.

\begin{figure}
\begin{center}
\includegraphics[width=1.\textwidth]{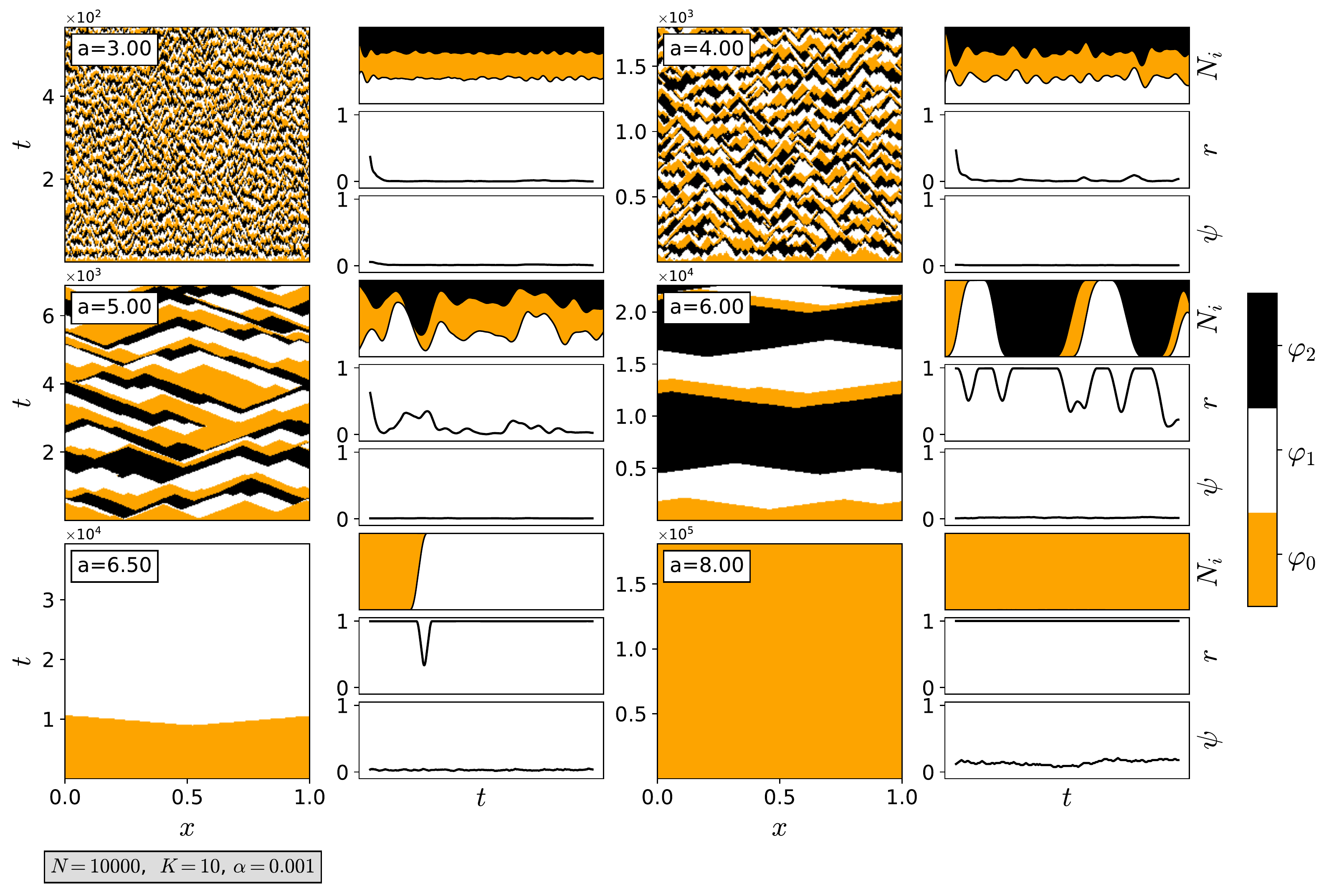}
\caption{\label{fig:trialpanel}
    Space-time plots, populations $N_i$, $\psi$ and $r$ order parameters. In
    all panels $N=10^4$, $K=10$ and $\alpha=10^{-3}$. The images on the left
    and center-right columns show space-time plots with position on the
    horizontal axis and time increasing upward.  Adjacent and to the right of
    each space-time plot, three graphs show the corresponding populations,
    $\psi$, and $r$ as functions of time.  Here we see that for a large system
    with low connectivity there are no regular oscillations. Instead, there are
    wave fronts that propagate and interfere and whose periods and amplitudes
    grow with $a$. 
    }
\end{center}
\end{figure}

Inspecting individual realizations of the dynamics in the small-$\alpha$ 
regime (Fig.~\ref{fig:trialpanel}) we see that the system never shows global
synchronization. It instead exhibits wave-like patterns that propagate in
both directions, similar to what is observed for large
negative coupling~\cite{escaff2014arrays}, but here for $a$ positive . The
amplitude and period of the wave increase with $a$ and with system size (for
fixed $\alpha$), which suggests there is an IP phase in the limit $N\to\infty$.

\subsection{Initial-configuration dependence}

\begin{figure}[b]
\begin{center}
\includegraphics[width=0.95\textwidth]{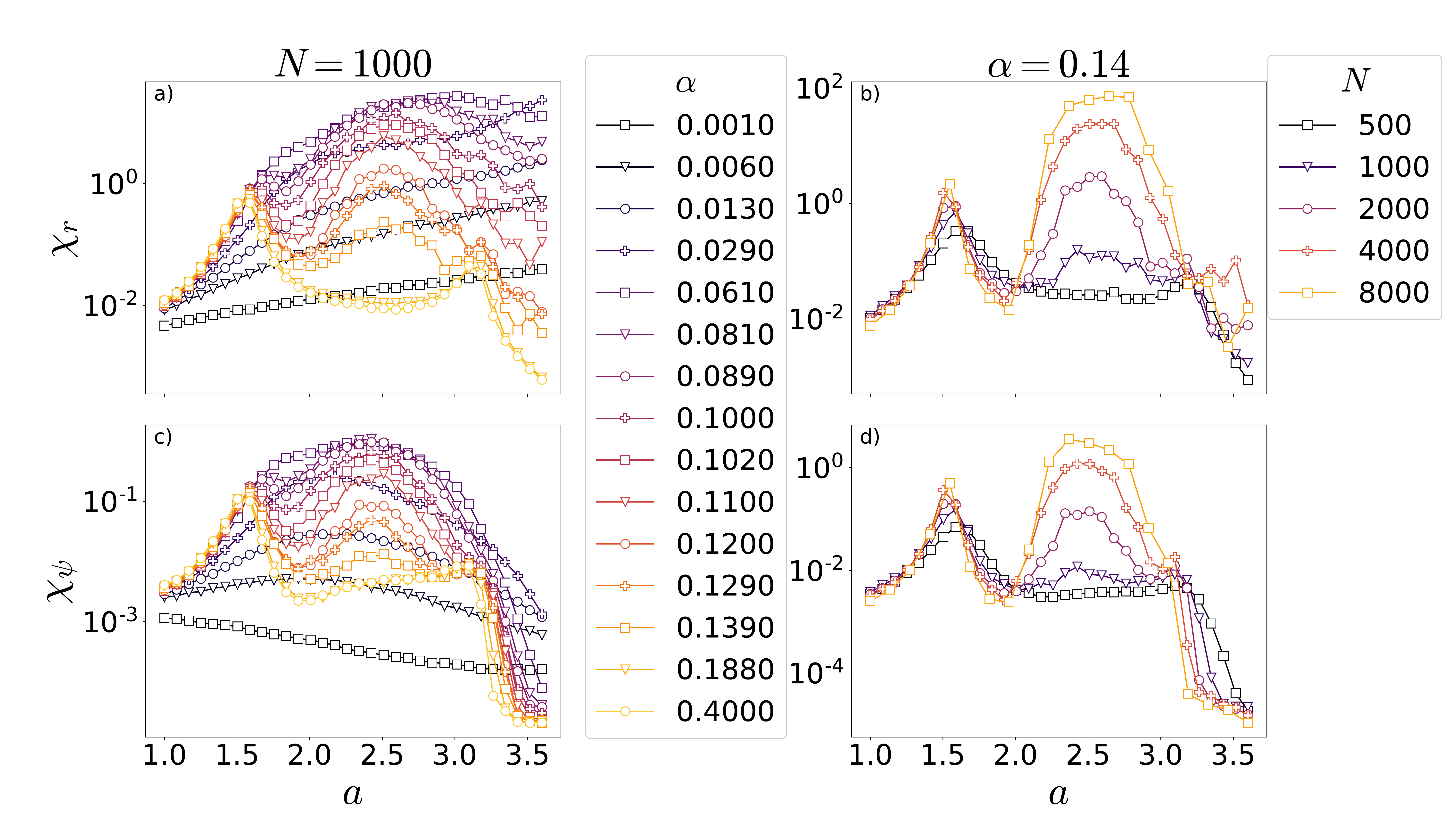}
\caption{\label{fig:chicurvesrandom}
    (Color online) Scaled variances $\chi_r$ and $\chi_{\psi}$ for random
    initial conditions.  Left column: fixed system size $N=1000$ with increasing
    $\alpha$.  Right column: fixed $\alpha=0.14$ and increasing system sizes.
    Points represent an average over 4000 independent realizations with random
    initial configurations.}
\end{center}
\end{figure}

Up to this point, all the results discussed were obtained using uniform initial
configurations (ICs).  In Fig.~\ref{fig:chicurvesrandom} shows $\chi_r$ and
$\chi_{\psi}$ for {\it random} ICs. The striking difference, compared to the
results for uniform ICs, is the presence of a middle peak between the two
identified previously. The scaling behaviors (panels \textbf{b} and \textbf{d})
suggest that the effect persists for large system sizes if $\alpha$ is held
constant. High values of $\chi$ result from multiple realizations of the
dynamics that produce net averages of the order parameter that differ from one
to another. At the (continuous) GO and IP phase transitions, the values of the
order parameter fluctuate strongly, giving rise to the peaks at $a=1.5$ and
$a\approx 3.1$. Another situation which may lead to high $\chi$ values is a
bistability between configurations that have large values of the order
parameter and others having a small one, even when fluctuations associated with
each configuration are small. This suggests a {\it discontinuous} phase
transition where for some values of $\alpha$ the system can relax to multiple
steady states. 

\begin{figure}
\begin{center}
\includegraphics[width=1.\textwidth]{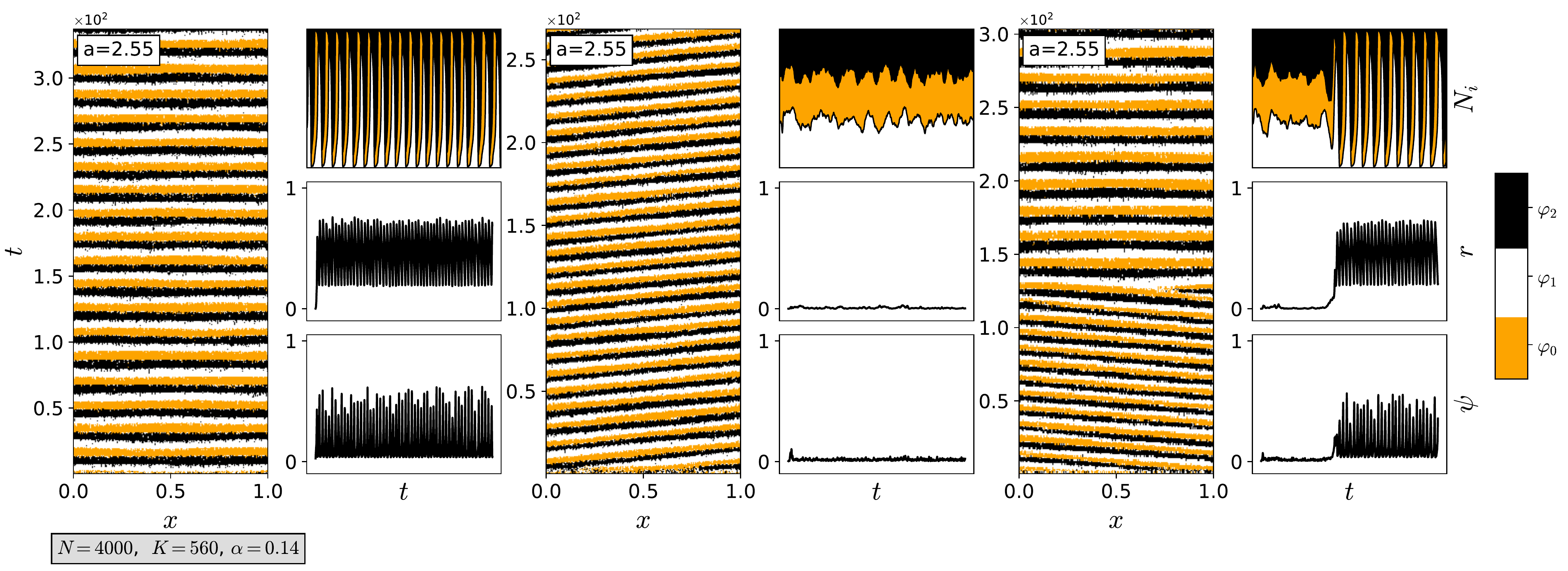}
\caption{\label{fig:trialpanel2}
    Single realizations with random initial configurations. Left: Global
    oscillations. Middle: travelling wave. Right: fluctuation-induced change
    from a travelling wave to global oscillations.  The fraction of realizations
    that converge to travelling waves is approximately $1.6\%$. In all panels
    $N=4000$, $K=560$, $a=2.55$.}
\end{center}
\end{figure}

Space-time plots of the dynamics with coupling $a\approx 2.5$ and $\alpha=0.14$
reveal that this is indeed the case (Fig.~\ref{fig:trialpanel2}).  Three
examples are shown in Fig.~\ref{fig:trialpanel2}: on the left is the familiar
globally synchronized state, while the middle panel shows a travelling-wave
state similar to those observed in \cite{escaff2014arrays}, (but here, for
large, positive coupling). The existence of a steady state with zero order
parameter but $a>a_c$ is surprising since Fig.~\ref{fig:chicurvesrandom}
suggests that wave-like solutions persist even in the limit $N\to\infty$ (with
fixed $\alpha$), where interaction ranges become infinite.

The rightmost panel in Fig.~\ref{fig:trialpanel2} shows a fluctuation-induced
change from a travelling wave to global synchrony.  Such transitions allow the
average order parameter to attain values between those associated with a wave
state and a globally synchronized one, being closer to one or the other
depending on what fraction of time it spent at that particular configuration.
Starting from {\it random} ICs, about 1.6\% of realizations exhibit travelling
waves, but since $\psi\approx 0$ for the waves and $\psi \sim {\cal O}(1)$ for
the globally synchronized case, the variances $\chi_r$ and $\chi_{\psi}$ are
sensitive even to small rates of occurrence. Due to fluctuations, waves do not
persist in smaller systems; sizes $N \geq 700$ are required. Smaller systems
exhibit either disordered phases with domains that increase in size and
duration as the coupling grows, or global synchrony if the interaction  range
$K$ is large enough.  In all cases, waves with exactly one spatial period over
the system are observed, while for negative coupling, multiple stable wave
numbers are found depending on the coupling magnitude \cite{escaff2014arrays}.

\begin{figure}[b]
\begin{center}
\includegraphics[width=1.\textwidth]{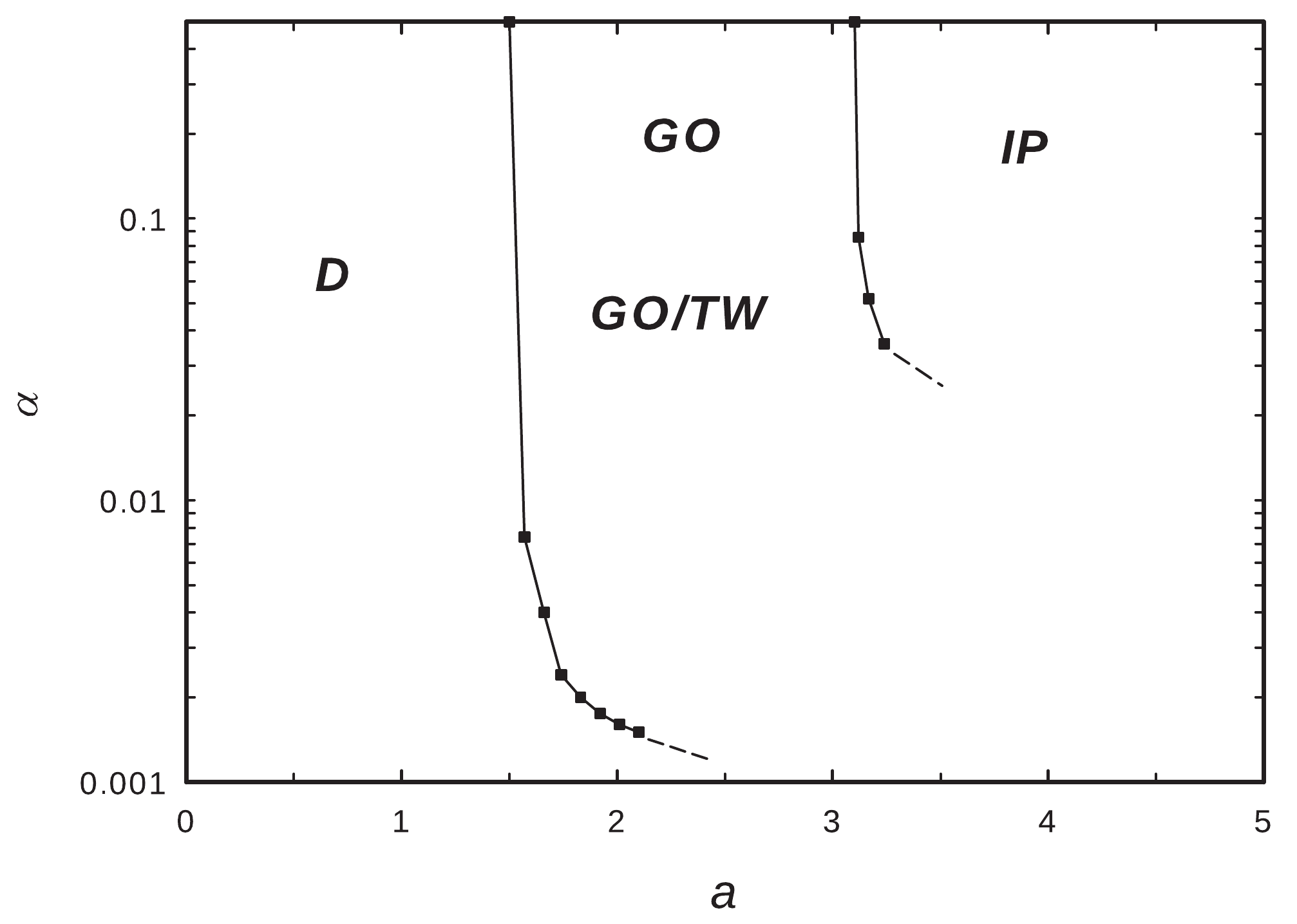}
\caption{\label{fig:phase_diagram}
    Regular rings: Preliminary phase diagram in the $a$-$\alpha$ plane.  Points
    are simulation results and (for $\alpha=0.5$) exact values.  Lines are
    guides to the eye.  Dashed lines represent our conjectures for how the
    phase boundaries continue.  Phases: disordered (D); global oscillation
    (GO); travelling-wave (TW); infinite- period (IP).}
\end{center}
\end{figure}

The existence of travelling waves can be understood by noting that, for
$\alpha<0.5$, the system consists of $G=N/2K$ domains. $G$ represents the
average path length for regular rings. (See Appendix~\ref{appendix:LC}.) When
$G$ is a multiple of three, the system is capable of containing a full
wavelength without oscillators in the center of the domains. The wavefronts are
then able to propagate as nucleation fronts \cite{assis2011infinite} giving
rise to travelling waves. In our simulations, with $N \leq10^4$, we find only
one stable wave-number (a wave with period $N$). This is in contrast with the
wave patterns observed in \cite{escaff2014arrays}, where multiple wave-numbers
were identified as stable depending on the magnitude of the coupling strength.

Although travelling waves are rare starting from random ICs, initializing with
solid blocks of 0s, 1s and 2s, each occupying one third of the system, the
ensuing evolution consists of a stable travelling wave in nearly all cases.
Thus the rarity for random ICs simply reflects the low probability of provoking
a wave, and does not reflect an intrinsic instability of the travelling-wave
state.  ICs with smaller blocks, such that the system contains two or more
waves, invariably yield, following a transient, a travelling wave whose
wavelength equals the system size.

With the known phase transition points for $\alpha=0.5$ and the data from
Figs.~\ref{fig:chicurvesrandom} and \ref{fig:opsplit} (for $N \leq 10^4$) we
are able to sketch the phase diagram (Fig.~\ref{fig:phase_diagram}). While the
phase boundaries are rather insensitive to changes in the connectivity for
relatively large $\alpha$ values, they veer to larger couplings for small
$\alpha$.


\section{\label{smallworld} The WCM on Small-World Networks}

\begin{figure}[b]
    \centering
    \begin{minipage}[t]{0.33\textwidth}
        \centering
        \includegraphics[width=0.9\textwidth]{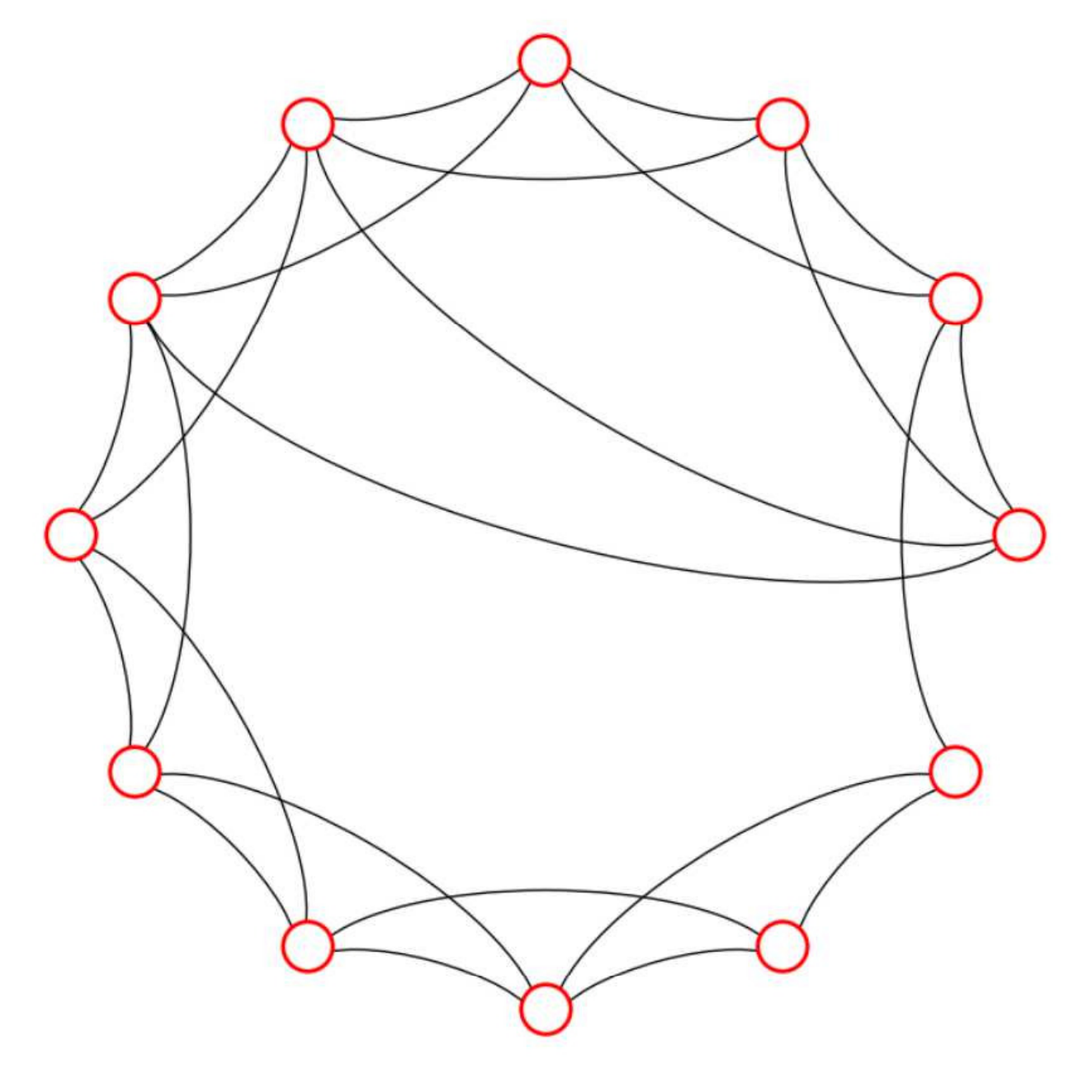}
        \caption{\label{fig:rewiredring}
        Example of a rewired regular ring with $N=12$, $K=2$ and $p=0.1$.
        }
    \end{minipage}
    \hfill
    \begin{minipage}[t]{0.62\textwidth}
        \centering
        \includegraphics[width=0.9\textwidth]{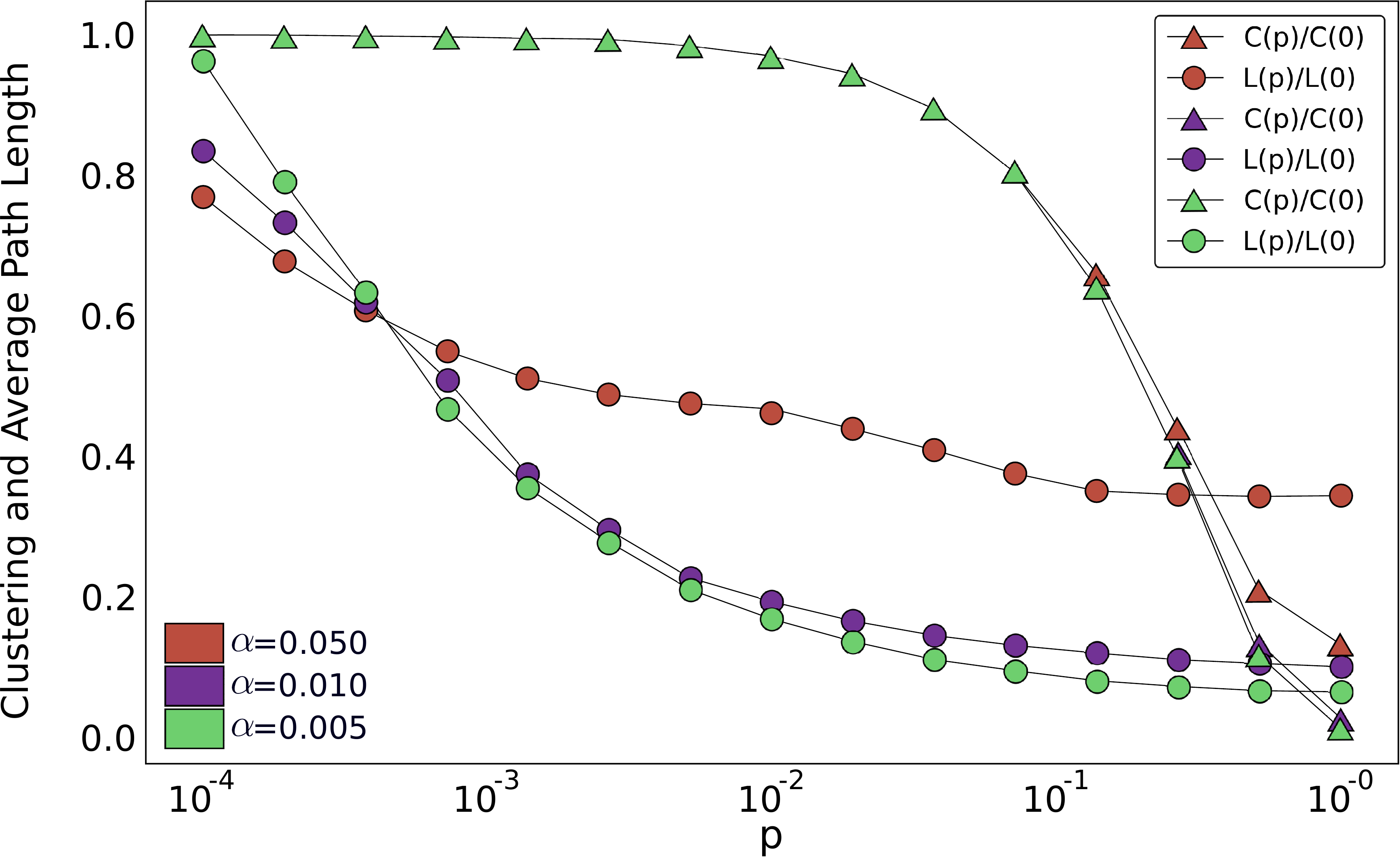}
        \caption{\label{fig:small-world}
            (Color online) Clustering coefficient $C$ and mean path length $L$
            versus rewiring probability $p$ for networks of $N=5000$ nodes.
            $K=250, 50, 25$ for red, blue, green curves respectively. The data
            represent averages over 400 independently generated networks.
        }
    \end{minipage}
\end{figure}

Regular rings can be used as the starting point for constructing {\it
small-world networks}, which are characterized by a small degree of separation
between nodes while maintaining local regions tightly clustered. A well known
algorithm for generating this type of network from regular rings was introduced
by Watts and Strogatz~\cite{watts1998collective}. Starting from a regular ring,
for each edge in the graph, the clockwise node of that edge is swapped with
probability $p$ for another randomly selected node, forbidding self-connections
or repeated edges. Some care may be taken to avoid disconnected graphs as a
result of this process, but this is unlikely for most parameter triplets ($N$,
$K$, $p$) considered here~\citep{watts1998collective}. An example of such a
rewired ring is shown in Fig.~\ref{fig:rewiredring}.

To characterize a network as small-world, we define two quantities:

\begin{itemize}
    \item {\it Mean path length} $L$: For nodes $i$ and $j$ in network ${\cal
        G}$, let $L_{ij}$ be the number of edges in the shortest path connecting
        these nodes.  Then the average path length of ${\cal G}$ is
        $L=\left<L_{ij} \right>$, where the average is over all pairs $(i,j)$
        with $i<j$.
    \item {\it Clustering coefficient} $C$: If node $i$ has $n_i$ neighbors,
        then the maximum possible number of connections among its neighbors is
        $m_i=\frac{n_i(n_i-1)}{2}$. Let $m^*_i$ be the number of connections
        among the neighbors of node $i$ in network ${\cal G}$. Then the
        clustering coefficient of ${\cal G}$ is $C=\left< \frac{m^*_i}{m_i}
        \right>$, where the average is over all nodes $i$.
\end{itemize}

Evidently, the maximum possible value for $C$ is $C=1$ and the minimum possible
value for $L$ is $L=1$. For networks generated by rewiring a regular ring graph,
$C$ and $L$ are functions of the ring-graph parameters $N$ and $K$, as well as
the rewiring probability $p$: $L\equiv L(N,K,p)$, $C\equiv C(N,K,p)$. As shown
in the Appendix, for regular rings (i.e., $p=0$), we have:

\begin{equation}
C(N,K,0)=
\begin{cases}
    0 ,                                      & \text{if } K<2 \\
    \frac{3K-3}{4K-2},                       & \text{if } 2 \leq K \leq
    \frac{N-1}{3} \\
    \frac{12K^2+6K-6KN+N^2-3N+2}{4K^2-2K},   & \text{if } \frac{N-1}{3} < K <
    \frac{N}{2} \\
    1,                                       & \text{otherwise}
\end{cases}
\end{equation}

\begin{equation}
L(N,K,0)=
\begin{cases}
    \frac{KG(G-1) + rG}{N-1}, & \text{if } K \leq \frac{N}{2} \\
    1,                        & \text{otherwise},
\end{cases}
\end{equation}

\noindent where $G$ is the largest integer smaller than $\frac{N-1}{2K}$ or,
using the floor operator,

\begin{equation*}
    G = \floor*{\frac{N-1}{2K}} .
\end{equation*}

For nonzero values of $p$ we generate graphs and take the averages for $C$ and
$L$, as shown in Fig.~\ref{fig:small-world}. For rewiring probabilities $p \in
\left( 0.001, 0.1 \right)$, rewiring preserves the clustering property while
greatly reducing the average path length, thus characterizing small-world
networks.

\begin{figure}
\begin{center}
\includegraphics[width=1.\textwidth]{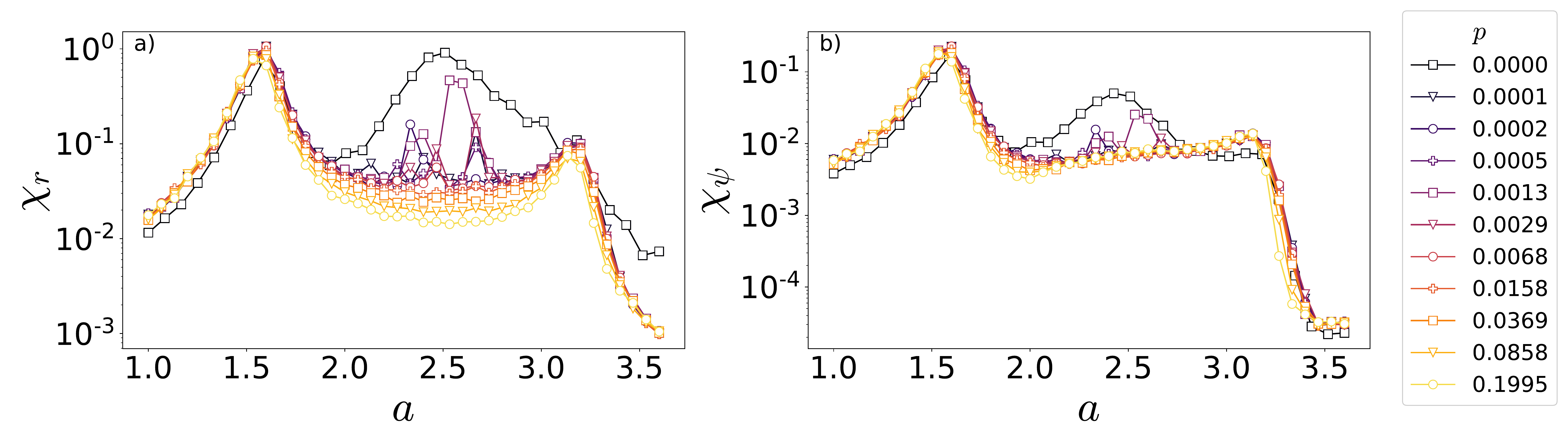}
    \caption{\label{fig:chicurvespvalue}
        (Color online) Small-world networks: Scaled variances, $\chi_r$ and
        $\chi_{\psi}$, for increasing rewiring probabilities. $N=1000$, $K=129$,
        $\alpha=0.129$ with random initial conditions. For $p \approx 0.01$ the
        curves follow their complete graph counterparts, even though the number
        of connections is half as many, and travelling waves become unstable.
    }
\end{center}
\end{figure}

\begin{figure}
\begin{center}
\includegraphics[width=0.8\textwidth]{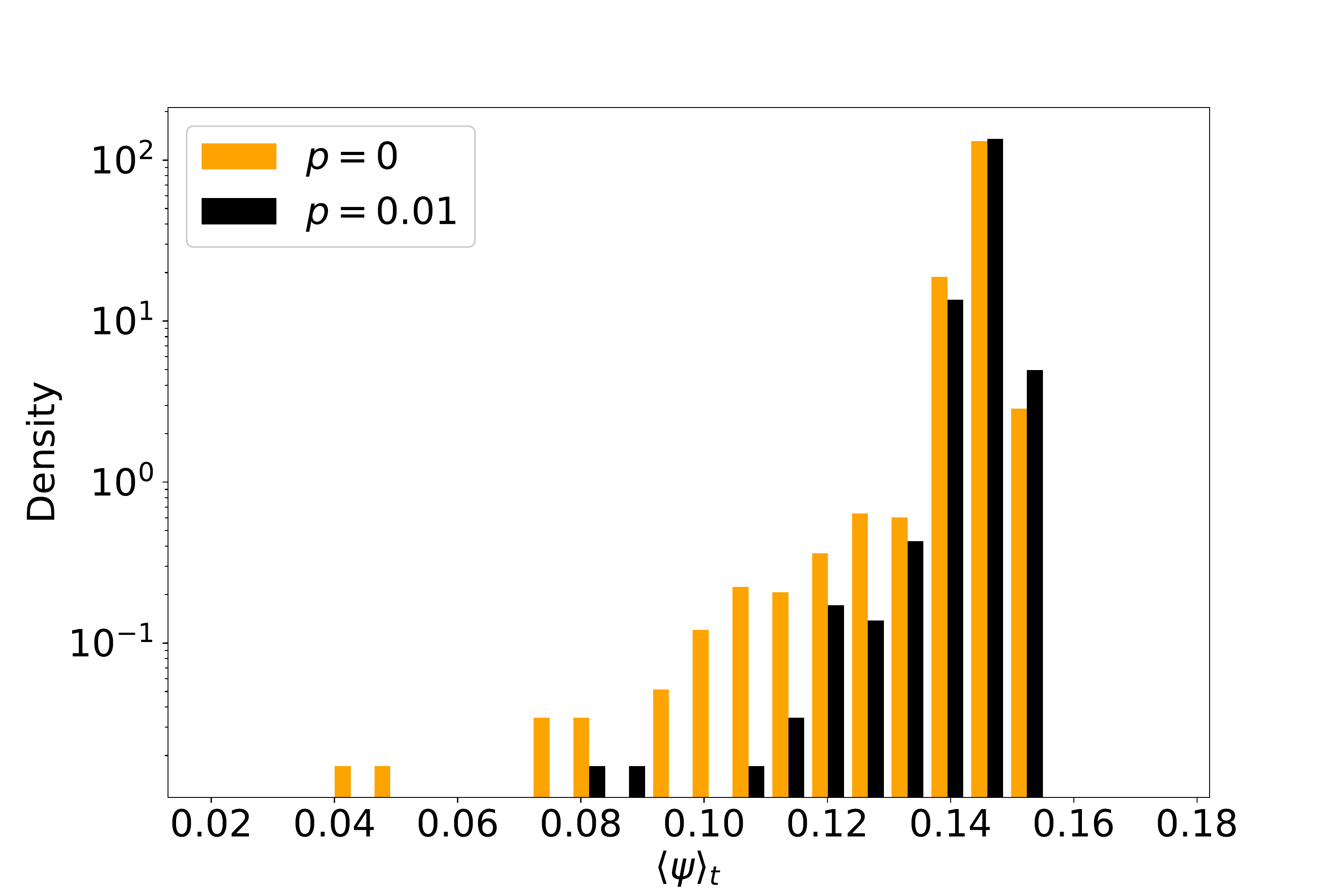}
    \caption{\label{fig:histogram}
        (Color online) Networks with $N=1000$, $K=129$ and coupling $a=2.5$.
        Histograms of $\langle \psi \rangle_t$ for 9000 realizations without
        rewiring (yellow), and 9000 realizations with rewiring probability
        $p=0.01$ (black).  The reduced frequency of small order-parameter
        values for $p=0.01$ compared with $p=0$ is evidence that rewiring
        destabilizes travelling waves, which are characterized by small values
        of $\psi$.}
\end{center}
\end{figure}

\begin{figure}
\begin{center}
\includegraphics[width=1.0\textwidth]{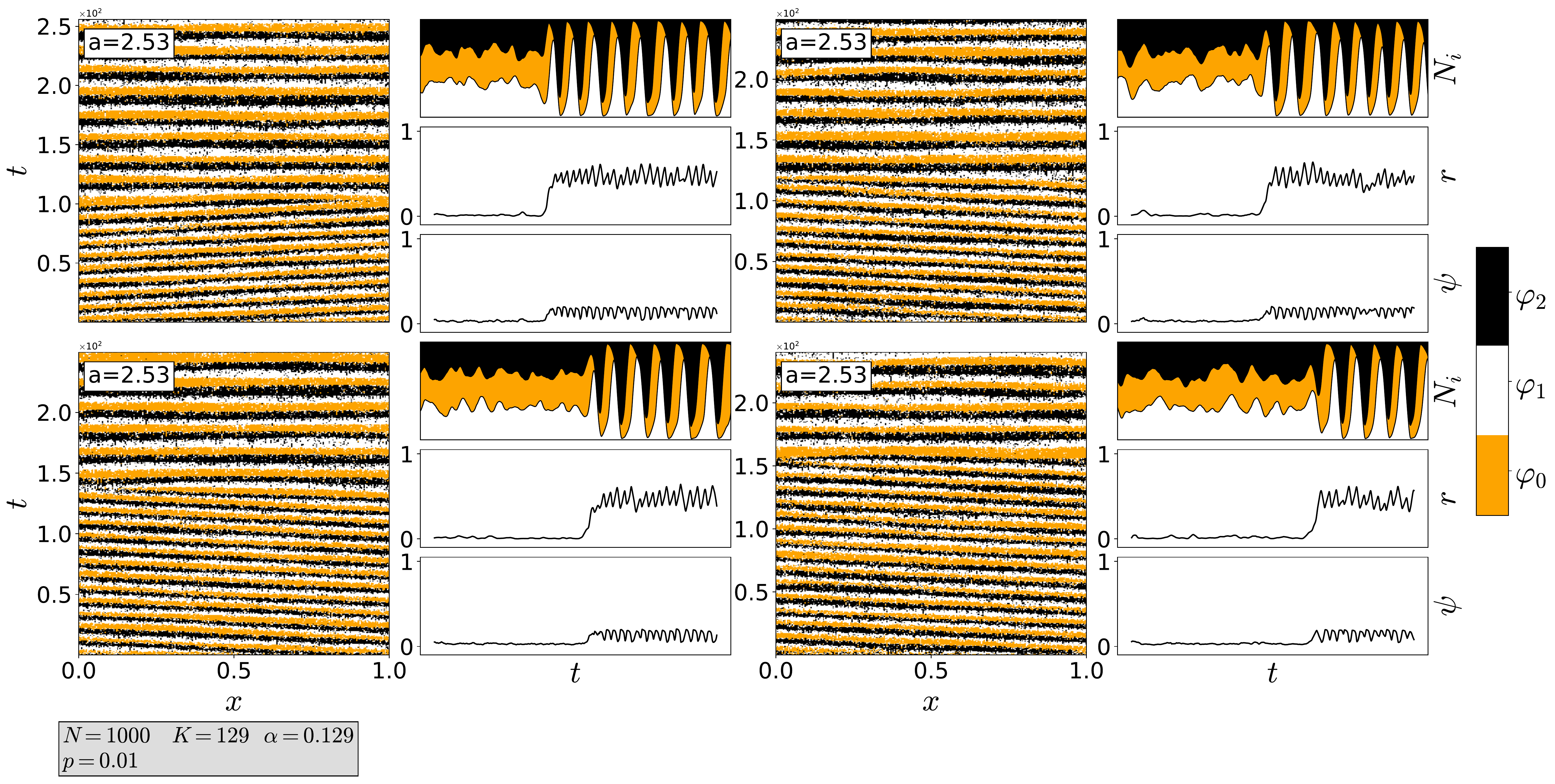}
\caption{\label{fig:trialpanel3}
    Wave states for $N=1000$, $K=129$, $p=0.01$ and $a=2.53$. Only $0.04\%$ of
    realizations (measured from 9000 realizations) with {\it random} initial
    configurations display traveling waves, which are also much more
    short-lived when compared to the $p=0$ equivalent system.}
\end{center}
\end{figure}

Since rewiring creates long-range interactions that reduce path lengths
globally, we expect it to facilitate synchronization. This is shown to be the
case in Fig.~\ref{fig:chicurvespvalue}, where $p$ is gradually increased:
realizations starting from {\it random} initial conditions are shown to readily
synchronize for very small values of $p$, leading to the usual three phases
identified for the complete graph. This means that the introduction of very few
global connections is sufficient to destabilize travelling waves; they are
virtually absent for $p \approx 0.01$. In Fig.~\ref{fig:histogram} a histogram
of the average order parameter per realization, $\left< \psi \right>_t$, shows
that the fraction of time spent in wave configurations is drastically reduced by
the introduction of the rewiring procedure. When $p\geq0.01$ and $a_c<a<a^c$,
the system quickly converges to the  GO phase even if it initially acquired a
wave-like solution (see Fig.~\ref{fig:trialpanel3}). Moreover, if system size is
increased at constant $\alpha$, smaller values of $p$ are sufficient to cause
the same effect, which is consistent with the fact that the number of long range
connections is proportional to $NK$.

\begin{figure}
\begin{center}
\includegraphics[width=0.9\textwidth]{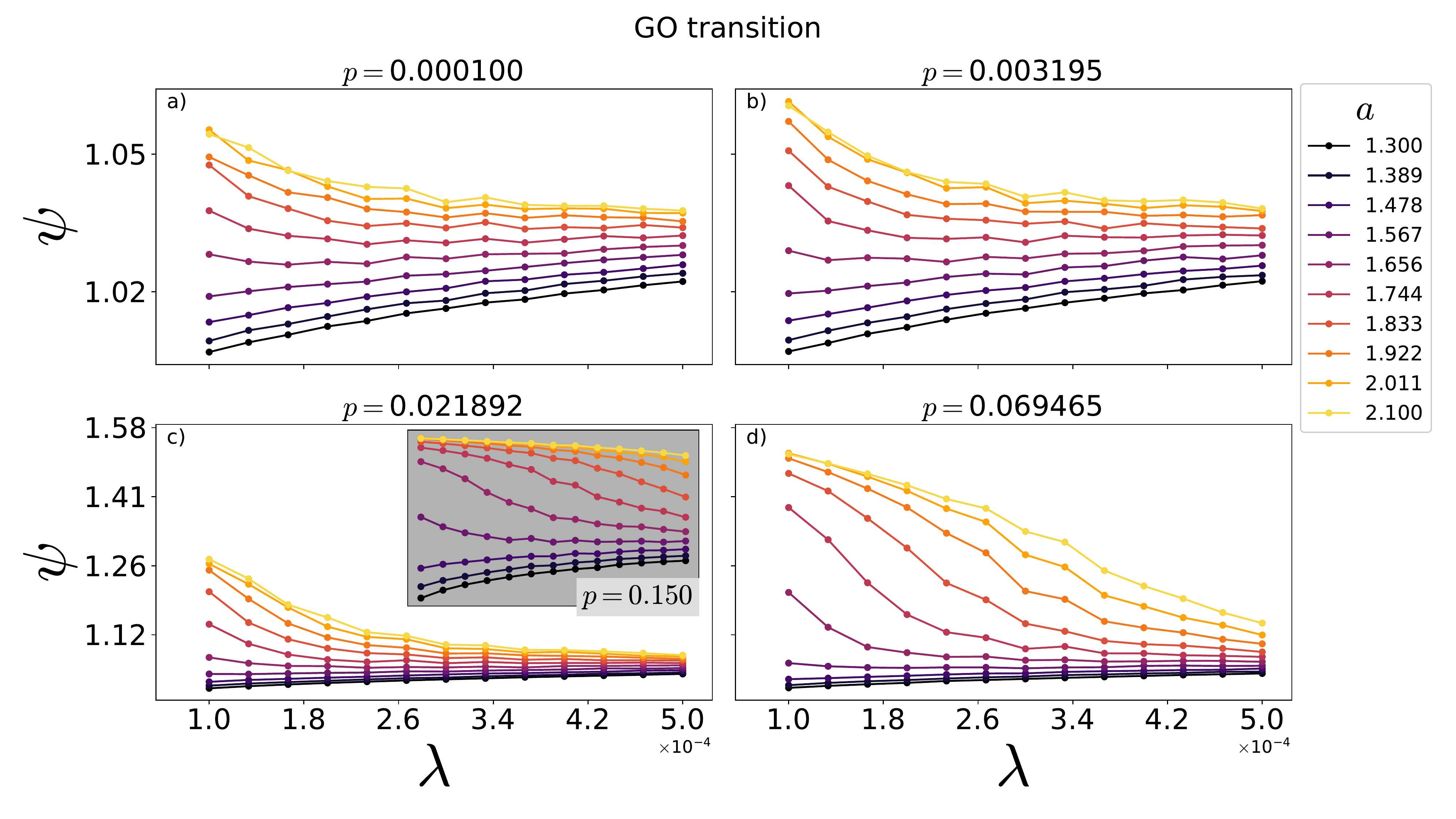}
\includegraphics[width=0.9\textwidth]{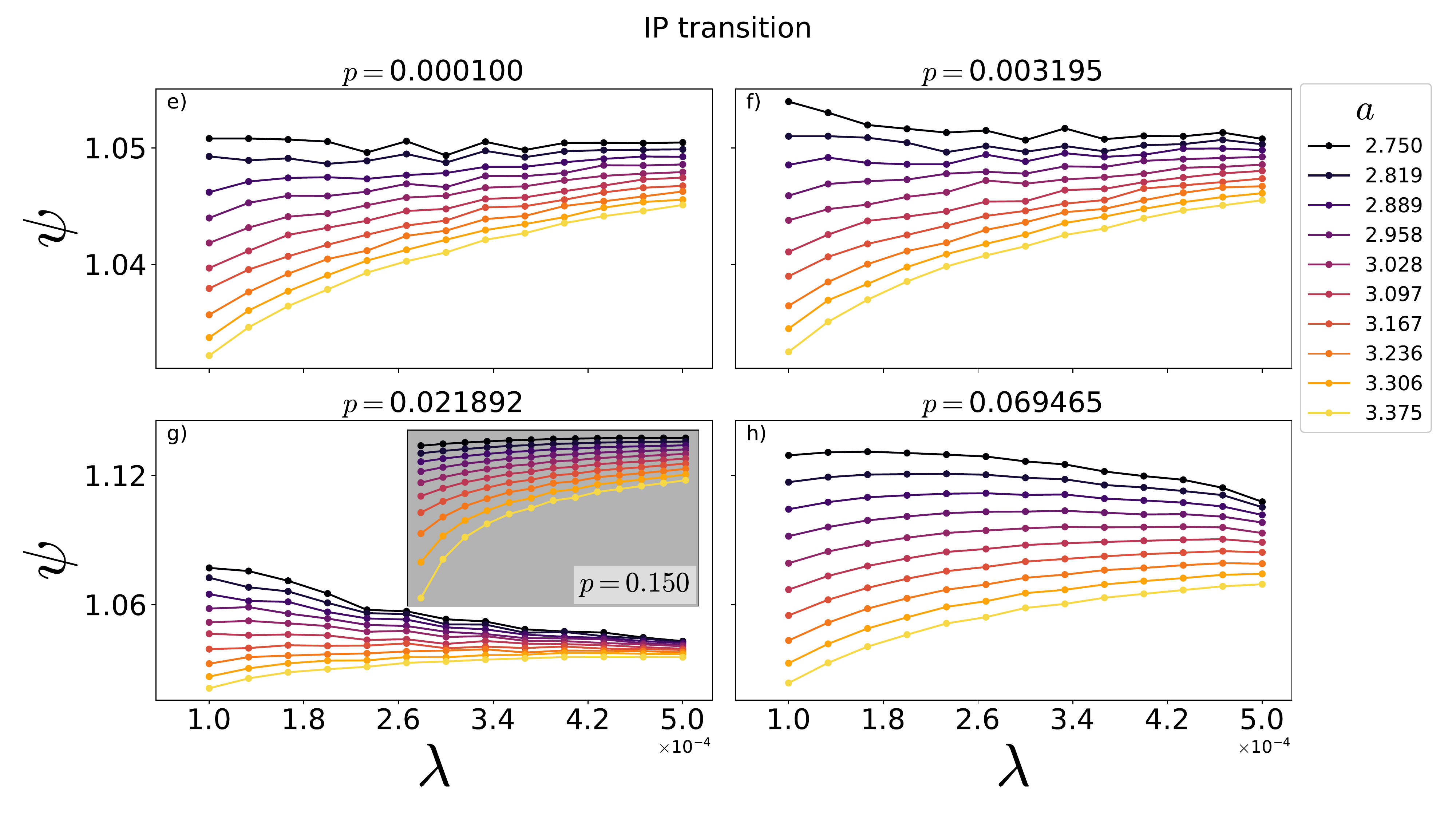}
\caption{\label{fig:pvalue}
    (Color online) Order parameter $\psi$ versus inverse system size $\lambda$
    for $\alpha=0.00524$ and various values of rewiring probability $p$ and
    couplings around the GO and IP transitions. Insets show the behavior for
    $p=0.15$, the highest rewiring probability used.}
\end{center}
\end{figure}

Following the same procedure as before, we plot in Fig.~ \ref{fig:pvalue} the
order parameters versus inverse system size $\lambda$. Here we fix $\alpha$ at
a low value ($\alpha=0.0052$) and vary $p$.  \footnote{{See full animations of
        figure \ref{fig:pvalue}:
\href{https://youtu.be/3iHXrDUwbqs}{GO transition}},
\href{https://youtu.be/J9OHw3DycAQ}{IP transition}}
Comparing these results with panels d) and h) of Fig.~\ref{fig:opsplit} we see
that the mean absolute value of $\psi$ is much greater when $p$ lies in the
small-world region, indicating greater synchrony among oscillators. Space-time
plots for values of $p>0$ show that indeed travelling waves become unstable, so
that the system exhibits only the three phases observed on the complete graph,
namely, disordered, GO and IP phases. The phase diagram remains similar to the
case of regular rings, displaying low sensibility to $\alpha$ except for very
small values where the discreteness of finite systems becomes apparent. The main
difference is that for any positive $p$ wave-like steady states are absent and
thus the phase diagram contains only three macroscopically distinct phases.

\clearpage

\section{\label{conclusions}Conclusions}

We study the WCM model, a discrete-phase oscillator model exhibiting global
synchronization and an infinite-period phase, on regular ring lattices and
small-world networks.  Each oscillator is coupled to a nonzero fraction
$\alpha$ of the others, but the connectivity is in generally much smaller than
for a complete graph.  Our results support the conjecture that all three phases
- disordered, globally synchronized, and infinite-period - appear for any
nonzero connectivity $\alpha$. 

Surprisingly, travelling waves also appear in the small-$\alpha$ regime on
regular ring lattices;  such waves may represent a long-lived metastable state.
(A previous study identified waves in the anti-coupling case
\cite{escaff2014arrays}.) In our studies travelling waves constitute only about
$1.6\%$ of the steady states reached from starting from {\it random} initial
configurations, and are prone to decay into global oscillations due to
fluctuations when system size is small.  The introduction of long-range
interactions through rewiring (i.e., using the Watts-Strogatz algorithm) can
lead to synchronization without increasing the total number of connections.
Rewiring also suppresses travelling waves by introducing long range
interactions.

The fact that regular rings are capable of sustaining travelling waves for
$a>a_c$ is surprising, showing that networks of oscillators might fail to
synchronize even in the presence of nonlocal interactions and strong coupling,
conditions which are sufficient for the synchronization on hypercubic lattices
of dimensions greater than 2 as well as on the complete graph. This raises the
possibility that WMCs admit wave-like steady states on cubic lattices,
similar to oscillations observed in the two-dimensional Belousov–Zhabotinsky
reaction.

Several other questions remain open for future study, for example, whether
travelling waves represent a stable phase for some range of parameters, or are
always metastable, and whether waves of wavelength smaller than the system size
$N$ are possible.  We have sketched a phase diagram for ring lattices, but
detailed information for the small-connectivity regime is lacking.  Since the
model exhibits a pair of continuous phase transitions, it is of interest to
develop a full scaling picture, including the effect of ordering fields
conjugate to the order parameters.  Finally the question of what simple
external  perturbations are capable of destroying synchronization or the
symmetry-broken phase may be of some practical interest.

\section{Appendix}
\subsection{\label{appendix:LC}Path Length and Clustering of Regular Ring
Lattices}

\subsubsection{Average Path Length}

The distance $L_{ij}$ between two nodes $i$ and $j$ is the minimum number of
edges that must be traversed to connect them. The average path length is defined
as the average distance between every possible pair of nodes in the network. For
a network with $N$ nodes this is:

\begin{equation}
    L = \frac{2}{N^2-N}\sum_{i<j}L_{ij}
\end{equation}

Consider the lowest node in a ring graph with $N$ nodes and $2K$ neighbors per
node. Going counterclockwise (CCW), there are $K$ nodes at distance $1$, then
$K$ nodes at distance $2$ and so on until we reach some region near the top. In
total, there will be $G$ groups of nodes, each with $K$ nodes, at distances
$1,2,3...,G$ from our starting point at the bottom. $G$ is given by the largest
integer smaller than $(N-1)/(2K)$. This can be written with the floor operation:

\begin{equation}
    G=\floor*{\frac{N-1}{2K}}
\end{equation}

This same procedure can be performed from the clockwise (CW) direction. Thus,
there are $2K$ nodes at distance $1$ and so on up to distance $G$. The last
group of nodes at the top is therefore at a distance $G+1$, but it contains less
than $2K$ nodes. Indeed, it contains a number $R$ of nodes equal to the
remainder of the integer division of $N-1$ by $2K$:

\begin{equation}
    R = N - 1 - 2KG
    \label{eq:R}
\end{equation}

\begin{figure}
\centering
\includegraphics[width=0.6\textwidth]{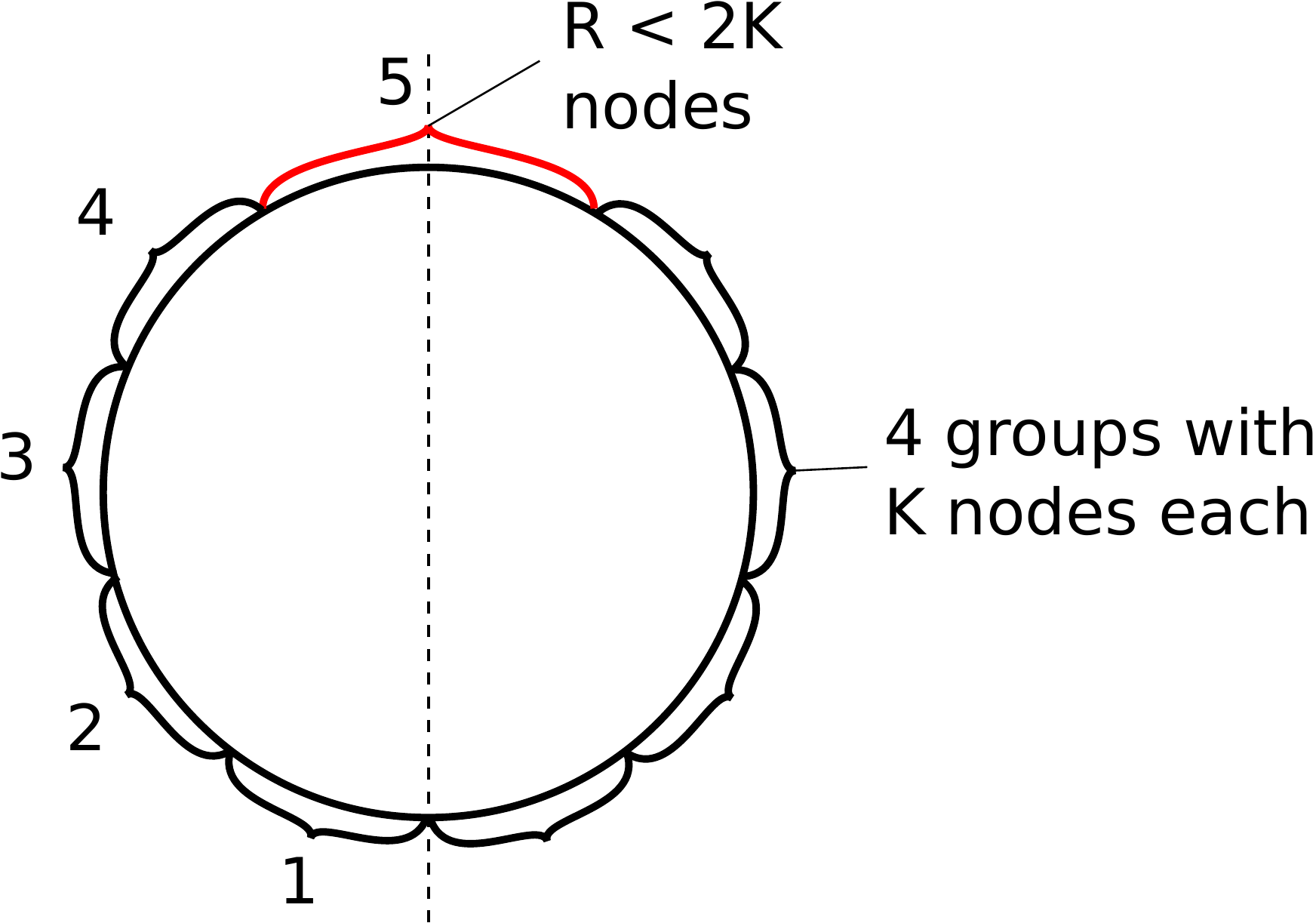}
\caption{Counting the number of nodes at distances visually. Here $G=4$ groups
    at distances $1,2,3,4$ from the bottom node.}
\label{fig:Lgroups}
\end{figure}

\noindent This reasoning can be visualized in Fig.~\ref{fig:Lgroups}.

Since there are $N-1$ pairs between the bottom node and all other nodes in the
lattice, the average distance $L_0$ between the bottom node and all other nodes
is then given by:

\begin{align}
    L_0 &= \frac{1}{N-1} \left[ 2K \sum_{i=1}^Gi + R(G+1) \right] \notag \\[9pt]
        &= \frac{1}{N-1} \left[ KG(G+1) + R(G+1) \right] \notag \\[9pt]
    L_0 &= \left(G + 1\right) \left( 1 - \frac{KG}{N-1} \right)
\end{align}

\noindent where we used equation \ref{eq:R} to substitute in for $R$.  Because
we started with an arbitrary node at the bottom, this result is true for any
given node in a regular ring, and thus we conclude that the average path length
for the whole network is just $L_0$.

\begin{align}
    L(N,K) = \left(G + 1\right) \left( 1 - \frac{KG}{N-1} \right) \notag \\[9pt]
    \text{with} \qquad G=\floor*{\frac{N-1}{2K}}
\end{align}

\subsubsection{Average Clustering}

The clustering coefficient for a node $i$ in the graph is defined as: let $n_i$
be the number of neighbors of some node $i$. Then, there are at most
$(n_i^2-n_i)/2$ connections between any two of its neighbors. Let $m_i$ be the
number of actual connections that are present in a particular graph. Then, the
clustering coefficient $C_i$ of node $i$ is given by:

\begin{equation}
    C_i = \frac{2m_i}{n_i^2-n_i}
\end{equation}

\noindent If there are $N$ nodes in the graph, the average clustering
coefficient is thus given by:

\begin{equation}
    C = \frac{1}{N}\sum^N_{i=1} C_i
\end{equation}

First, consider a ``close-up" of a section of a regular ring where $N\gg K$.
Consider a node $i$ with $K$ CW neighbors and $K$ CCW neighbors. Let $m$ be the
number of connections between neighbors of node $i$. We can manually count $m$
for some values of $K$:

\begin{figure}
\centering
\includegraphics[width=0.75\textwidth]{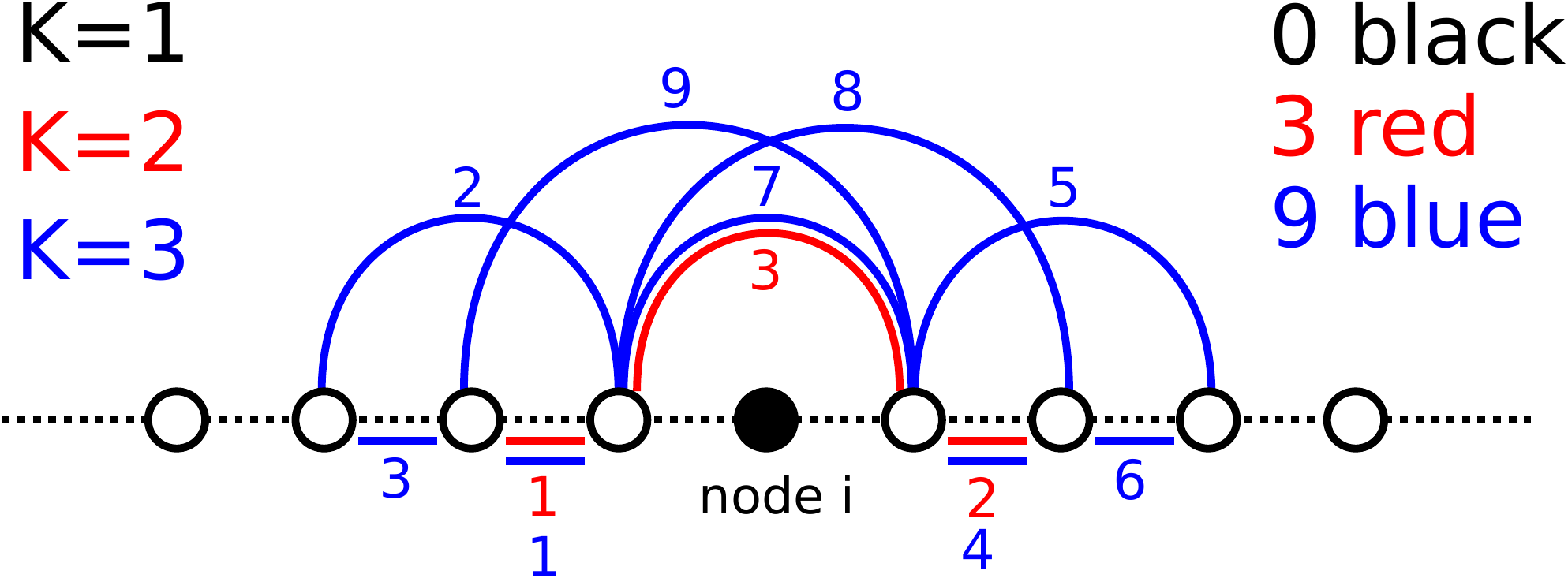}
\caption{
    (Color online) Counting $m$ for $k=1,2,3$. We note three contributions to
    $m$: a fully connected group of $K$ CW neighbors (left), a fully connected
    group of $K$ CCW neighbors (right) and connections that go ``over" the
    center node.
    }
\label{fig:clustering}
\end{figure}

\begin{align*}
    m &= 0 \qquad \text{if} \qquad K \leq 1\\
    m &= 3 \qquad \text{if} \qquad K = 2\\
    m &= 9 \qquad \text{if} \qquad K = 3
\end{align*}

The general case for $N \gg K$ can be counted by summing the contributions of
the three groups as shown in figure \ref{fig:clustering}. Two fully connected
groups of $K$ nodes each contributes with $(K^2-K)$ connections. The connections
that bypass node $i$ also contribute with $(K^2-K)/2$ connections and thus we
have.

\begin{align}
    m &= \frac{3}{2}(K^2-K)
    \label{eq:m}
\end{align}

\noindent Now we divide equation \ref{eq:m} by the total number of connections
between the $2K$ neighbors of node $i$ to get its clustering coefficient $C_i$:

\begin{align}
    C_i(N,K) = \frac{3K-3}{4K-2}
    \label{eq:ci}
\end{align}

\begin{figure}
\centering
\includegraphics[width=0.9\textwidth]{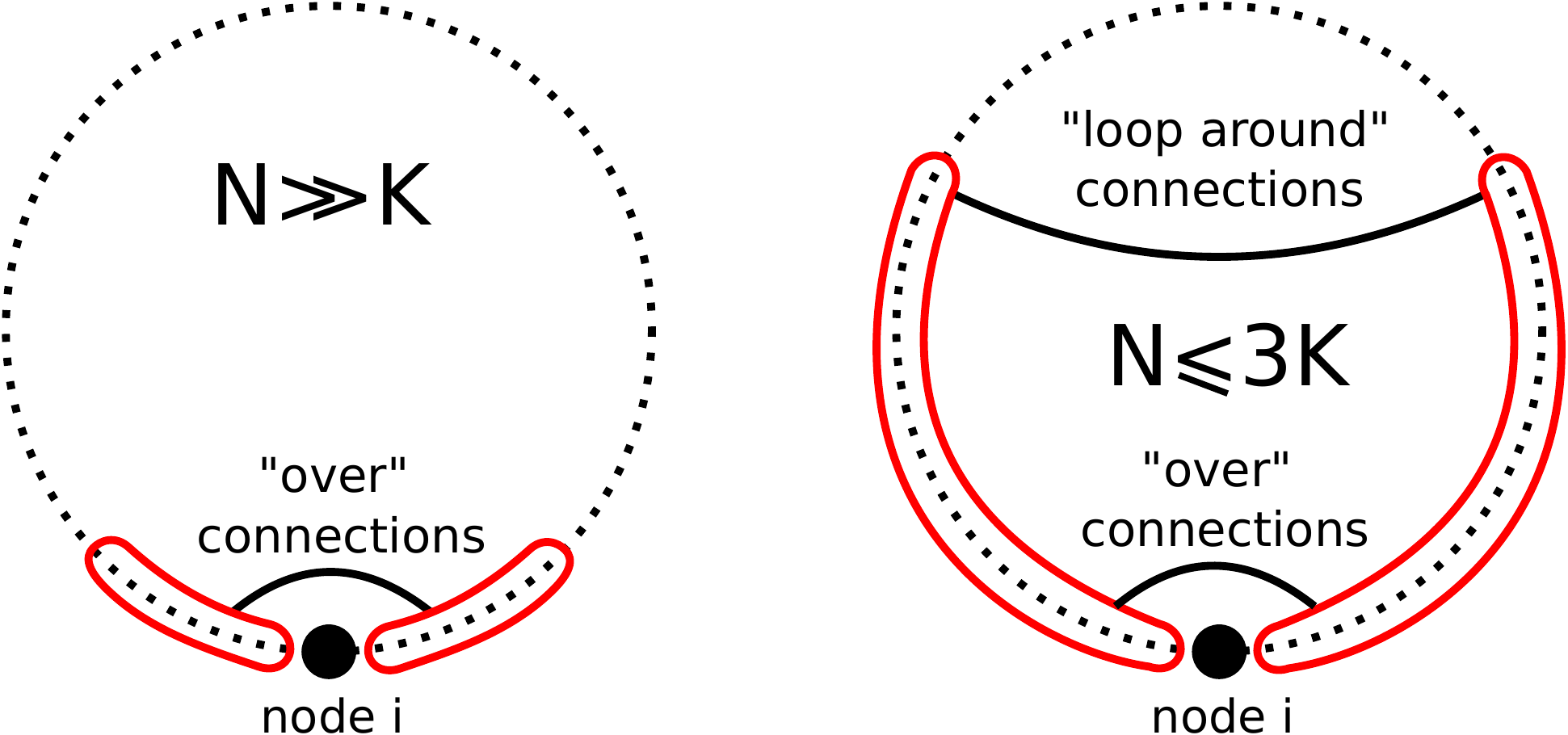}
\caption{
    Connections that contribute to the clustering coefficient of node $i$.  The
    red regions represent the CW and CCW groups of neighbors of node $i$, and
    they are fully connected each within itself. Additional connections are made
    going ``over" node $i$, but when $N\leq3K$ there are additional connections
    that loop around the opposite side of the lattice.
    }
\label{fig:loop}
\end{figure}

\noindent When $N$ is not so large compared to $K$, additional connections
between the CCW and CW neighbors may appear by looping around the opposite side
of node $i$, as depicted in figure \ref{fig:loop}. These additional connections
will be present whenever the remaining nodes that are neither in the CW or CCW
group are fewer than $K$. The number of such nodes is just $N-2K-1$, which gives
us the condition $N\leq 3K$ for additional connections to be present.

The number of ``loop around" connections that will be present will depend on how
many nodes there are in the remaining group after removing node $i$ and its
immediate neighbors. Let this number be denoted by $D=N-2K-1$. Then, the number
of additional connections will be given by

\begin{align}
    (K-D) + (K-D-1) + ... + 2 + 1 = \frac{(K-D+1)(K-D)}{2}
    \label{eq:looparound}
\end{align}

Adding equation \ref{eq:looparound} to \ref{eq:m} we get

\begin{align}
    m = \frac{3}{2}(K^2 - K) + \frac{1}{2}(3K-N+1)(3K-N+2)
\end{align}

And thus the clustering now becomes:

\begin{align}
    C_i(N,K) = \frac{3K-3}{4K-2} + \frac{(3K-N+1)(3K-N+2)}{4K^2-2K}
    \label{eq:ci1}
\end{align}

This formula holds up to the point when $D=0$ or $N=2K+1$. For all values $D\leq
0$ the regular ring is in fact a complete graph, where every node connects to
every other. In these cases the clustering coefficient is always equal to one.
Formulas \ref{eq:ci} and \ref{eq:ci1} together offer a complete expression for
the clustering of node $i$. Since $C_i=C_j \quad \forall i,j$, this is just the
average clustering of the whole network and we finally get:

\begin{align}
    C(N,K) =
    \begin{cases}
        \frac{3K-3}{4K-2} \quad &\text{if} \quad N>3K \\[9pt]
        \frac{3K-3}{4K-2} + \frac{(3K-N+1)(3K-N+2)}{4K^2-2K} \quad &\text{if}
        \quad 3K \geq N > 2K+1 \\[9pt]
        1 \quad &\text{else}
    \end{cases}
    \label{eq:ci2}
\end{align}

\begin{acknowledgments}
     KLR acknowledges the financial support from CNPq and CAPES, Brazil. RD
     acknowledges support of CNPq under project 303766/2016-6.
\end{acknowledgments}

\clearpage
\bibliography{references.bib}

\end{document}